\documentclass[useAMs,usenatbib,a4paper]{mn2e}
\usepackage{savesym}
\usepackage{graphicx}
\expandafter\let\csname equation*\endcsname\relax
  \expandafter\let\csname endequation*\endcsname\relax 
\usepackage{subfig}
\usepackage{times}
\usepackage{amsmath}
\usepackage{amssymb}
\usepackage{verbatim}
\usepackage{array}
\usepackage{times}
\usepackage[total={17.8cm,24.0cm},centering]{geometry}

\newcommand{\W}{{\rm \,W}}
\newcommand{\pc}{{\rm \,pc}}
\newcommand{\kpc}{{\rm \,kpc}}
\newcommand{\be}{\begin{equation}}
\newcommand{\ee}{\end{equation}}
\newcommand{\eea}{\end{eqnarray}}
\newcommand{\bea}{\begin{eqnarray}}

\newcommand{\m}{\mathrm}

\title[An accelerating jet model based on observations of M87]{Synchrotron and inverse-Compton emission from blazar jets - II. An accelerating jet model with a geometry set by observations of M87}
\author[William J. Potter and Garret Cotter]{William J. Potter\thanks{E-mail:
will.potter@astro.ox.ac.uk (WJP)} and Garret Cotter
\\
Oxford Astrophysics. Denys Wilkinson Building, Keble Road, Oxford, OX1 3RH, United Kingdom}
\begin{document}

\date{}

\pagerange{\pageref{firstpage}--\pageref{lastpage}} \pubyear{2011}

\maketitle

\label{firstpage}

\begin{abstract}
In this paper we develop the jet model of Potter \& Cotter (2012) to include a magnetically dominated accelerating parabolic base transitioning to a slowly decelerating conical jet with a geometry set by recent radio observations of M87.  We conserve relativistic energy-momentum and particle number along the jet and calculate the observed synchrotron emission from the jet by calculating the integrated line of sight synchrotron opacity through the jet in the rest frame of each section of plasma.  We calculate the inverse-Compton emission from synchrotron, CMB, accretion disc, starlight, broad line region, dusty torus and narrow line region photons by transforming into the rest frame of the plasma along the jet.  

We fit our model to simultaneous multi-wavelength observations of the Compton-dominant FSRQ type blazar PKS0227-369, with a jet geometry set by M87 and an accelerating bulk Lorentz factor consistent with simulations and theory.  We investigate models in which the jet comes into equipartition at different distances along the jet and equipartition is maintained via the conversion of jet bulk kinetic energy into particle acceleration.  We find that the jet must still be magnetically dominated within the BLR and cannot be in equipartition due to the severe radiative energy  losses.  The model fits the observations, including radio data, very well if the jet comes into equipartition outside the BLR within the dusty torus (1.5pc) or at further distances (34pc).  The fits require a high power jet with a large bulk Lorentz factor observed close to the line of sight, consistent with our expectations for a Compton-dominant blazar.  We find that our fit in which the jet comes into equipartition furthest along the jet, which has a jet with the geometry of M87 scaled linearly with black hole mass, has an inferred black hole mass close to previous estimates.  This implies that the jet of PKS0227 might be well described by the same jet geometry as M87. 

\end{abstract}


\section{Introduction}

Blazars are thought to be active galactic nuclei (AGN) with relativistic jets close to our line of sight.  The non-thermal synchrotron and inverse-Compton emission from the jet are Doppler-boosted so that blazars are the most luminous and variable of AGN, with apparent superluminal motion visible in their jets.  The boosted jet emission dominates the spectrum of blazars with synchrotron emission extending from radio to UV/x-rays and inverse-Compton emission from x-rays to very high energy $\gamma$-rays.  Blazars are observed to produce radio emission which is flat or nearly flat in flux. 

Radio observations show that AGN jets are approximately continuous axisymmetric plasma jets.  Recent observations show that on large scales jets are approximately conical in geometry and measurements of the frequency dependent core-shift of M87 and other AGN favour a conical jet with conserved magnetic energy.  VLBI observations of M87 by \cite{2011Natur.477..185H} show the blunt base of the jet is within $14-23R_{S}$ of the central black hole.  Observations of the geometry of the jet in M87 by \cite{2011arXiv1110.1793A} over a large range in length scales show that the jet is in fact parabolic close to the base and becomes conical at $10^{5}R_{S}$.  These observations are supported qualitatively by simulations of AGN jets powered by the Blandford-Znajek mechanism (see \cite{2006MNRAS.368.1561M}) which show an accelerating parabolic base which becomes conical near to the point where the plasma reaches equipartition.

The detailed mechanism behind the production of a relativistic plasma jet from a black hole is not yet understood fully.  It is thought that the energy source of the jet comes from the rotation of the central black hole and that magnetic field from the accretion disc is twisted by the rotation of the black hole creating a Poynting flux (\cite{1977MNRAS.179..433B}).  We expect from simulations and theory that jets start as Poynting dominated (magnetic field dominated) and that the particles in the jet are accelerated to a large bulk Lorentz factor by the magnetic field until the jet plasma approaches equipartition where it reaches a terminal bulk Lorentz factor.  The accelerating region is thought to be parabolic as indicated by observations \cite{2011arXiv1110.1793A}, simulations (\cite{2006MNRAS.368.1561M}) and analytic work (\cite{2006MNRAS.367..375B}), whilst the outer regions of the jet are consistent with being approximately conical (\cite{2011arXiv1110.1793A}, \cite{2011Natur.477..185H}, \cite{2011arXiv1103.6032S} and \cite{Krichbaum:2006pw}) with a slowly decelerating bulk Lorentz factor (\cite{2009MNRAS.398.1989M} and \cite{2010ApJ...710..743D}).   

In this paper we develop our conical jet model from \cite{2012MNRAS.423..756P} (hereafter Paper I) to include an accelerating parabolic base which transitions to a slowly decelerating conical jet after $10^{5}R_{S}$, motivated by the recent radio observations of M87 and consistent with simulations and theory.  Our model conserves relativistic energy-momentum and particle number throughout the jet and includes energy losses to the particles (electrons and positrons, hereafter referred to as electrons) via radiative and adiabatic losses.  We integrate the observed line of sight synchrotron opacity through the jet by transforming into the plasma rest frame.  We also include the effect of inverse-Compton scattering external photons from the CMB, an accretion disc, starlight, accretion disc photons scattered by the broad line region (BLR) and accretion disc photons reprocessed by a dusty torus and narrow line region (NLR).  We calculate the inverse-Compton emission from these external sources by Lorentz transforming into the rest frame of the plasma.  This is the first time that an inhomogeneous jet model has included an accelerating parabolic base transitioning to a decelerating conical jet.  In this paper we wish to develop a realistic, physically rigorous jet model motivated by and consistent with observations and simulations.

A successful jet model should be able to reproduce the simultaneous multi-wavelength observations of blazars across all wavelengths including radio observations, which are usually neglected.  To test whether our model is capable of producing realistic spectra we fit it to the simultaneous multi-wavelength spectrum of the Compton-dominant FSRQ PKS0227-369 from \cite{2010ApJ...716...30A}.  

Currently the most popular model of blazars are time-dependent fixed radius one-zone models e.g. \cite{1998A&A...333..452K}, \cite{2002ApJ...581..127B}, \cite{2000ApJ...536..729L} and \cite{2007A&A...476.1151T}.  These models consider the injection of an arbitrary electron population into a spherical blob with a constant bulk Lorentz factor and calculate the time-dependent emission from synchrotron, SSC and EC mechanisms.  One-zone models best describe flaring events observed in the jets, however, they struggle to reproduce radio observations.  Investigations into the effects of deceleration in one-zone models include those by \cite{1998NewA....3..157D} and \cite{2009ApJ...692.1374B} where plasmoids are decelerated by collisions which accelerate particles.  It is important for a realistic model of blazars to be compatible with radio observations of the morphology, core-shift and flat radio spectrum, which are incompatible with one-zone models.  

Investigations of extended inhomogeneous jets include those conducted by \cite{1979ApJ...232...34B}, \cite{1980ApJ...235..386M}, \cite{1981ApJ...243..700K}, \cite{1982ApJ...256...13R}, \cite{1984ApJ...285..571M} and \cite{1985A&A...146..204G}, \cite{1985ApJ...298..114M}, \cite{2000A&A...356..975K}, \cite{2000AAS...197.8401M}, \cite{2004ApJ...613..725S}, \cite{2008ApJ...689...68G}, \cite{2009ApJ...699.1919P}, \cite{2010MNRAS.401..394J} and \cite{2011arXiv1112.2560V}.  Early investigations of conical jets were motivated by the original work by \cite{1979ApJ...232...34B} which showed that a conical jet which conserves magnetic energy produces a flat radio spectrum. 

There have been a number of investigations into accelerating or decelerating jets including those by \cite{1989ApJ...340..181G}, \cite{1998ApJ...506..621G}, \cite{2001MNRAS.325.1559S}, \cite{2003ApJ...589L...5G} and \cite{2008MNRAS.390L..73B}.  These investigations have included various models for acceleration and deceleration in parabolic, conical and cylindrical jets, however, no previous work has modelled both the accelerating parabolic base and decelerating conical sections of the jet or calculated the line of sight synchrotron opacity through the jet.  Most investigations have variously treated some aspects of the jet physics and emission thoroughly whilst neglecting others.  The early, pioneering work by \cite{1989ApJ...340..181G} is the only previous investigation to include an accelerating parabolic base transitioning to a conical jet.  However, this investigation did not consider electron energy losses, inverse-Compton scattering of external photons, energy-momentum conservation along the jet or a decelerating conical section. In light of recent observations and simulations we feel it is valuable to develop a physically motivated model of jets which includes a thorough treatment of the relevant physics and which we can use to fit to blazar spectra in order to try to answer some of the fundamental questions about the blazar population. 

In this paper we will first explain the basic properties of the jet including its geometry and variable bulk Lorentz factor.  We then show how energy-momentum and particle number are conserved along the jet.  We calculate the observed synchrotron emission and inverse-Compton emission from external photons of electrons in the jet.  We then show the results of fitting the model to the Compton-dominant FSRQ type blazar PKS0227.  Finally, we discuss the physical properties and characteristics of our model fit.

\section{Jet Model}

We develop the conical jet model presented in Paper I.  In this paper we wish to extend the model to include a variable geometry, a variable bulk Lorentz factor, inverse-Compton scattering of external photons and adiabatic energy losses.

\subsection{Geometry}

We choose our model to have a parabolic base which transitions to a conical jet at a lab frame distance $x_{\m{con}}$ from the black hole.  The radius of the jet $R(x)$ is given by

\be
R(x)=R_{0}+A_{\m{par}}x^{B_{\m{par}}} \qquad \m{for} \,\,\,\, x\leq x_{\m{con}},
\ee
\be
R(x)=R_{0}+A_{\m{par}}x_{\m{con}}^{B_{\m{par}}} +(x-x_{con})\tan(\theta_{\m{opening}}) \,\,\, \m{for} \,\,\,\, x>x_{con}
\ee

where $x$ is the lab frame jet length and $\theta_{\m{opening}}$ is the lab frame half opening angle of the jet.  We fit the geometry of our model to the observed geometry of M87 (\cite{2011arXiv1110.1793A}), though we leave the opening angle of the conical section of the jet as a free parameter since this is likely to vary with the bulk Lorentz factor of the jet as $\sim \frac{1}{\gamma_{\m{bulk}}}$.  We assume the geometry of our jet scales linearly with black hole mass.  This means that the entire geometry of our jet is described by only two free parameters (the others are set by the observations of M87) the black hole mass $M$ and the lab frame conical half-opening angle $\theta_{\m{opening}}$.  

\begin{figure}
	\centering
		\includegraphics[width=8 cm, clip=true, trim=1cm 1cm 0cm 1cm]{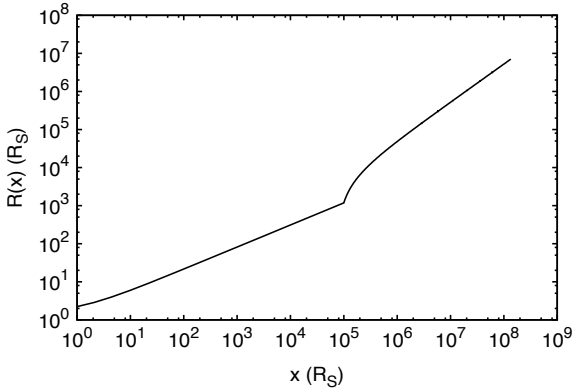}
			
	\caption{This figure shows the geometry of our jet as used in the model fit to PKS0227 where the dimensions are given in terms of the Schwarzschild radius of the black hole.  The jet starts parabolic with the geometry fixed to match observations of M87, the jet transitions to a conical geometry at $x_{\m{con}}=10^{5}R_{S}$ with lab frame half-opening angle $\theta_{\m{opening}}$.  }
	\label{fig2}
\end{figure}

We do not expect the geometry of the inner region of all AGN jets to be the same as that of M87, however, it is very interesting to see whether a jet with this geometry is compatible with the observations of a blazar.  Our model is flexible and so can incorporate any length-radius relation if new information on the geometry of other AGN jets becomes available in the future.  

\subsection{Acceleration and deceleration}

We set the bulk Lorentz factor of the jet plasma to be a function of $x$.  We use the relationship between the bulk Lorentz factor and the jet radius predicted by theory (\cite{2006MNRAS.367..375B}) and consistent with simulations (\cite{2006MNRAS.368.1561M})

\be
\gamma_{\m{bulk}}(x) \propto x^{1/2}.
\ee

Setting the bulk Lorentz factor at the base of the jet to be $\gamma_{0}$ and the maximum Lorentz factor at the end of the accelerating parabolic region to be $\gamma_{\m{max}}$, for $x \leq x_{\m{con}}$ the bulk Lorentz factor is given by

\be
\gamma{_\m{bulk}}(x)=\gamma_{0}+\left(\frac{\gamma_{\m{max}}-\gamma_{0}}{x_{\m{con}}^{1/2}}\right)x^{1/2}  \,\,\,\,\, \m{for} \,\,\,x \leq x_{\m{con}}.
\ee

Once the jet has reached equipartition it can no longer significantly accelerate the plasma by converting magnetic energy into bulk kinetic energy and so the jet stops accelerating and becomes conical.  The jet now slowly decelerates (\cite{2002MNRAS.336..328L}, \cite{2005MNRAS.358..843H}, \cite{2009MNRAS.398.1989M} and \cite{2010ApJ...710..743D}) due to interaction with its environment and Compton-drag.  In this model we will assume the bulk Lorentz factor decreases with $\log(x)$.  The conical section of the jet starts with a bulk Lorentz factor $\gamma_{\m{max}}$ and decelerates by the end of the jet at $x=L$ to a Lorentz factor of $\gamma_{\m{min}}$.  Under these assumptions the bulk Lorentz factor is given by

\be
\gamma_{\m{bulk}}(x)=\gamma_{\m{max}}-\left(\frac{\gamma_{\m{max}}-\gamma_{\m{min}}}{\log \left(\frac{L}{x_{\m{con}}}\right)}\right) \log \left(\frac{x}{x_{\m{con}}}\right), \,\,\,\,\,\, x \geq x_{\m{con}}. \label{dec}
\ee

The dependence of the deceleration of a jet with jet length is not well known so we have assumed this logarithmic decrease.  Our model is flexible so we can easily change the dependence of the bulk Lorentz factor to observations of deceleration if they become available.   

\subsection{Energy and particle conservation for the plasma}

In our model the jet plasma starts as magnetically dominated at the base, this magnetic energy is converted to bulk kinetic energy of the plasma as the jet is accelerated in the parabolic region by a Poynting flux (\cite{2006MNRAS.368.1561M} and \cite{2009MNRAS.394.1182K}).  We do not attempt to model the physical mechanism behind the plasma acceleration since in this investigation we are interested primarily in the observed emission of the jet which depends only on the physical properties of the plasma.  

We treat the plasma as being an isotropic perfect fluid in its instantaneous rest frame along the whole jet.  This corresponds to the magnetic field being randomly oriented on small scales and isotropic and the particles being relativistic with an isotropic velocity distribution in the rest frame of the plasma.  This is an idealised situation since it requires the magnetic field and electrons to isotropise after a section of plasma has been accelerated or decelerated, whilst this is the highest entropy state for the plasma we assume the timescale over which the plasma isotropises is short compared to the acceleration timescale and this may not be valid.  However, this assumption of a plasma in which both the magnetic field and particle distribution are isotropic and homogeneous in its rest frame is used in almost all models of jet emission. 

The magnetic field close to the base of the jet where the plasma is Poynting dominated is expected to be approximately parabolic (\cite{2006MNRAS.368.1561M}) and so the electromagnetic stress-energy tensor is not that of a perfect fluid, however, in a model of jet emission we are only interested in the regions closer to equipartition where high energy particles exist or are accelerated.  In these regions close to or downstream of shocks we expect the magnetic field to be much more turbulent and so the approximation of an isotropic tangled field is appropriate for these jet regions which contribute significantly to the overall emission.  

We treat the field as tangled on small scales and particles as relativistic so that $P=\frac{\rho}{3}$, the energy-momentum tensor in the rest frame $T'^{\mu \nu}$ is given by

\be
T'^{\mu \nu}=\begin{pmatrix} \rho & 0 &0 &0 \\ 0 & \frac{\rho}{3} &0 &0\\ 0 &0 &\frac{\rho}{3} &0\\ 0 &0 &0 &\frac{\rho}{3}\end{pmatrix},
\ee

where $\rho$ is the total energy density in the plasma rest frame and $\rho=\rho_{e}+\rho_{B}$ where $\rho_{e}$ is the particle energy density and $\rho_{B}$ is the energy density of the magnetic field.  For a section of plasma which is travelling with a bulk Lorentz factor $\gamma_{\m{bulk}}(x)$ in the lab frame its energy-momentum tensor in the lab frame $T^{\mu \nu}$ is given by

\be
T^{\mu \nu}(x)=\Lambda^{\mu}_{\,\, \m{a}} T'^{\m{a} \m{b} } \Lambda _{\m{\,\,b}}^{ \nu}=\begin{pmatrix} \frac{4}{3}\gamma_{\m{bulk}}(x)^{2} \rho & \frac{4}{3}\gamma_{\m{bulk}}(x)^{2} \rho &0 &0\\\frac{4}{3}\gamma_{\m{bulk}}(x)^{2} \rho &\frac{4}{3}\gamma_{\m{bulk}}(x)^{2} \rho &0 &0\\0&0&\frac{\rho}{3}&0\\0&0&0&\frac{\rho}{3}\end{pmatrix} \label{LTEM},
\ee

where $\Lambda^{\mu}_{\,\, \nu}$ is the Lorentz transformation tensor and we have assumed that the jet is always relativistic so that $\beta \approx 1$.  In order to satisfy the conservation of energy and momentum along the jet our energy-momentum tensor must satisfy $\nabla_{\mu}T^{\mu \nu}=0$.  In the lab frame our jet does not depend on time and quantities only have an $x$ dependence so all other derivatives vanish.  To conserve the total energy-momentum in a slice of width one metre as it propagates along the jet we also need to take into account the change in cross-sectional area of the jet (which is invariant under a Lorentz boost in the $x$-direction) as it expands. 

\be
\frac{\partial }{\partial \m{x}}(\pi R^{2}(x)T^{1 \nu}(x))=\frac{\partial}{\partial \m{x}}\left(\frac{4}{3}\gamma_{\m{bulk}}(x)^{2}\pi R^{2}(x)\rho(x)\right)=0. \label{ce}
\ee

We also need to conserve the number of existing energetic electrons in the jet as the bulk Lorentz factor of the jet changes, so our electron energy density must satisfy the continuity equation.

\be
\nabla_{\mu}J^{\mu}=\frac{\partial}{\partial \m{x}}(\pi R^{2}(x)\rho_{\m{e}}(x)U(x)^{1})=0,
\label{cc}
\ee

where $\rho_{\m{e}}$ is the electron energy density in the rest frame and $U(x)^{\mu}$ is the lab frame four velocity of the plasma.  These two conservation equations can be used to calculate the rest frame magnetic energy density and electron energy density as the jet accelerates and magnetic energy is converted into bulk kinetic energy.  We use the finite difference formulations of Equations \ref{ce} and \ref{cc} to evolve our magnetic field and electron population along the jet.  We do this by calculating the synchrotron and inverse-Compton emission of the initial electron population, $N(x,E_{e})$, in a section of lab frame width d$x$.  Our code calculates an appropriate width d$x$ adaptively such that the shortest radiative lifetime of the population is resolved and the radius of the jet does not change by more than $5\%$ over the width of the section.  We then calculate the resulting electron spectrum as a result of radiative and adiabatic losses over the section using Equations \ref{ad} and \ref{loss}.  In our code we also allow for the injection of electrons (with the same initial electron spectrum given by Equation \ref{ne}) produced from shock acceleration in the section, so our modified electron spectrum including energy losses and particle injection becomes $N'(x,E'_{e})$.  

\bea
N'(x,E'_{e})=N(x,E_{e})&-&\frac{P_{\m{tot}}(x,\m{d}x',E_{e})\times 1s}{cE_{e}} \nonumber\\
&&+A_{\m{inj}}(x,\m{d}x)E_{e}^{-\alpha} e^{-E_{e}/E_{\m{max}}},
\eea

where $E'_{e}$  is defined in Equation \ref{Ead} as the energy $E_{e}$ after adiabatic losses have been taken into account and $P_{\m{tot}}(x,\m{d}x',E_{e})$ is the total emitted power from the section of rest frame width d$x'$ from Equation \ref{loss}.  The rest frame magnetic energy injected into electrons in a slab of rest frame width 1m is $\pi R(x)^{2} \Delta \rho_{\m{inj}}(x)$, so $A_{\m{inj}}(x,\m{d}x)$ is determined via

\be
\pi R^{2}(x) \Delta \rho_{\m{inj}}(x)=\int_{E_{\m{min}}}^{\infty} A_{\m{inj}}(x,\m{d}x)E_{e}^{1-\alpha} e^{-E_{e}/E_{\m{max}}}\m{d}E_{e}.
\ee

In this investigation we will not inject electrons arbitrarily along the jet, so we take $A_{\m{inj}}(x,\m{d}x)=0$.  We wish to include the possibility of an arbitrary injection of electrons in these calculations, however, since this will allow us to investigate the effect of flaring sections of the jet in future work.  We use the continuity Equation \ref{cc} to calculate the overall normalisation of the modified electron spectrum in the rest frame of the next section starting at $x+\m{d}x$

\be
\m{Norm}=\frac{\gamma_{\m{bulk}}(x)R^{2}(x)}{\gamma_{\m{bulk}}(x+\m{d}x)R^{2}(x+\m{d}x)},
\ee

This takes into account the effect of differential length contraction on the number density of electrons as the population moves along the jet. The modified electron spectrum contained in one metre of the jet is given by

\be
N''(x+\m{d}x,E_{e}')=\m{Norm} \times N'(x,E'_{e}).
\ee

We now need to ensure conservation of the total energy-momentum along the jet.  In the accelerating part of the jet we determine the magnetic field in the section located at $x+\m{d}x$ using the finite difference formulation of Equation \ref{ce} 

\bea
&&\gamma_{\m{bulk}}(x)^{2}(\rho_{\m{B}}(x)+\rho'_{\m{e}}(x))\pi R^{2}(x)= \nonumber\\
&& \gamma_{\m{bulk}}(x+\m{d}x)^{2} (\rho_{\m{B}}(x+\m{d}x)+\rho''_{\m{e}}(x+\m{d}x))\pi R^{2}(x+\m{d}x), \nonumber \\
\label{cont}
\eea

where the energy densities are defined as

\bea
\rho'_{e}(x)&=&\frac{\int_{E_{\m{min}}}^{\infty} E'_{e}N'(x,E'_{e}) \m{d}E'_{e}}{\pi R^{2}(x)}, \qquad \rho_{B}(x)=\frac{B^{2}(x)}{2 \mu_{0}}, \nonumber \\ 
&&\rho''_{e}(x+\m{d}x)=\frac{\int_{E_{\m{min}}}^{\infty} E'_{e}N''(x+\m{d}x, E'_{e}) \m{d}E'_{e}}{\pi R^{2}(x+\m{d}x)},
\eea

In the accelerating part of the jet the initial electron distribution at the beginning of the section starting at $x+\m{d}x$ is given by

\be
N(x+\m{d}x,E_{e})=N''(x+\m{d}x,E_{e}).
\ee 

We assume that the plasma in the conical part of the jet is always in equipartition and the jet is ballistic so no longer suffers adiabatic losses as in Paper I.  In this case the bulk kinetic energy of the jet is converted into the acceleration of electrons at shocks and amplification of the magnetic field as the jet decelerates and so we use Equation \ref{ce} to determine the amount of energy injected into electrons through shock acceleration in a slab of rest frame width 1m $\Delta \rho_{\m{dec}}(x+\m{d}x)\pi R(x+\m{d}x)^{2}$.

\bea
&&\gamma_{\m{bulk}}(x)^{2}(\rho_{\m{B}}(x)+\rho'_{\m{e}}(x))\pi R^{2}(x)= \gamma_{\m{bulk}}(x+\m{d}x)^{2}... \nonumber\\
&& \times (\rho_{\m{B}}(x+\m{d}x)+\rho''_{\m{e}}(x+\m{d}x)+\Delta \rho_{\m{dec}}(x+\m{d}x))\pi R^{2}(x+\m{d}x), \nonumber \\
\label{cont1}
\eea

where we ensure the jet remains in equipartition in the conical section via the constraint

\be
\rho_{\m{B}}(x+\m{d}x)=\rho''_{\m{e}}(x+\m{d}x)+\Delta \rho_{\m{dec}}(x+\m{d}x).
\ee

The electron spectrum at $x+\m{d}x$ in the decelerating part of the jet is given by 

\be
N(x+\m{d}x,E_{e})=N''(x+\m{d}x,E_{e})+C(x+\m{d}x)E_{e}^{-\alpha} e^{-E_{e}/E_{\m{max}}},
\ee
\be
\Delta \rho_{\m{dec}}(x+\m{d}x)\pi R(x+\m{d}x)^{2}=C(x+\m{d}x)\int_{E_{min}}^{\infty} E_{e}^{1-\alpha} e^{-E_{e}/E_{\m{max}}}\m{d}E_{e}.
\ee

\section{Physical conditions at the base of the jet}

We wish to calculate the magnetic field strength and electron population in the rest frame in terms of the physical parameters of the jet.  For a relativistic jet with total kinetic luminosity of $W_{j}$, $\frac{W_{j}}{c}$ is equal to the total energy in a section of plasma of width one metre in the $x$ direction in the lab frame $E_{j}$.  

\be
E_{j}= T^{00}(x=0) \m{d}V= \frac{4}{3}\gamma_{0}^{2}\rho(x=0) \pi R_{0}^{2},
\ee
\be
\rho(x=0)=\frac{3E_{j}}{4\pi R_{o}^{2} \gamma_{0}^{2}},
\ee
\be
\rho(x=0)=\rho_{B}(x=0)+\rho_{e}(x=0), \qquad A_{\m{equi}}(x)=\frac{\rho_{B}(x)}{\rho_{e}(x)},
\ee
\be
\rho_{B}(x=0)=\frac{B_{0}^{2}}{2 \mu_{0}}, \qquad
\ee

We assume that the high energy electrons have been accelerated through shocks (\cite{1978MNRAS.182..147B}) or magnetic reconnection (\cite{2001ApJ...562L..63Z} and \cite{2007ApJ...670..702Z}).  Our code is flexible and allows us to inject an arbitrary electron population at different distances along the jet so we are not tied to a specific mechanism for particle acceleration.  In this paper we characterise the spectrum as being a power law with index $\alpha$ with an exponential cutoff at energy $E_{\m{max}}$ and with a minimum energy $E_{\m{min}}$.  A power law electron distribution with an exponential cutoff is an approximation to the steady state solution to the diffusion loss equation for a shock including radiative energy losses (see for example \cite{1985ApJ...288...32B}).  The number of electrons contained in a slab of width one metre in the rest frame of the plasma is given by

\be
N_{e}(x=0,E_{e})=A(x=0) E_{e}^{-\alpha} e^{-\frac{E_{e}}{E_{\m{max}}}}.
\label{ne}
\ee

In general our electron distribution will evolve due to energy losses and additional injection of electrons to take a more general form where $A(x,E_{e})$ becomes a function of electron energy and length along the jet.  Our initial value of electron energy density is therefore

\be
\rho_{e}(x=0)=\int ^{\infty}_{E_{\m{min}}} E_{e} \frac{N_{e}(x=0,E_{e})}{\pi R_{0}^{2}}  \m{d}E_{e}.
\ee

These equations specify the evolution of the plasma along the jet conserving energy-momentum and electron number density.  We now turn our attention to calculating the observed synchrotron emission from the jet.

\section{Synchrotron emission} 

In this section we shall generalise the calculation of synchrotron emission from Paper I to the case of a jet with a variable bulk Lorentz factor.  In the lab frame we wish to calculate the observed luminosity at a given observation angle of the jet.  To do this we will calculate the emission and absorption of synchrotron radiation corresponding to an observed lab frame frequency and observation angle as it propagates along the jet by Lorentz transforming into the rest frame at every section of the jet.  Observed synchrotron radiation at a frequency $\nu$ and angle $\theta_{\m{obs}}$ corresponds to an emitted frequency $\nu'$ and angle $\theta'_{\m{obs}}$ in the rest frame of a section of plasma with bulk Lorentz factor $\gamma_{\m{bulk}}$ given by a Lorentz transformation

\be
\nu'=\frac{\nu}{\delta_{\m{Doppler}}(x)}, \,\,\,\,\, \delta_{\m{Doppler}}(x)=\frac{1}{\gamma_{\m{bulk}}(x)(1-\beta(x) \cos(\theta_{\m{obs}}))},
\ee
\be
\theta'_{\m{obs}}=\cos^{-1}\left(\frac{\cos(\theta_{\m{obs}})-\beta(x)}{1-\beta(x) \cos(\theta_{\m{obs}})}\right), \,\,\,\,\, \gamma_{\m{bulk}}(x)=\frac{1}{\sqrt{1-\beta(x)^{2}}}.
\ee

We track photons emitted from interior sections of the jet as they move through outer sections by transforming to the rest frame of each section.  For a given observed lab frame frequency $\nu$ all the contributing synchrotron photons from different sections with different bulk Lorentz factors will have the same frequency $\nu'$ and emission angle $\theta_{\m{obs}}'$ in the rest frame of the section at $x$.  This means that to calculate the emission at an observed frequency $\nu$ we need to calculate the synchrotron emission at different frequencies $\nu'$ in sections of plasma with different bulk Lorentz factors.  To calculate the line of sight opacity to a section we need to integrate the opacity at rest frame frequency $\nu'$ across the rest frame path lengths of all the outer sections.  

We use the result from Paper I for the distance travelled through the plasma by a synchrotron photon in the rest frame  

\be
\m{d}r'=\gamma_{\m{bulk}} \m{d}x \left(\frac{1}{\cos(\theta_{\m{obs}})}-\beta(x)\right),
\ee

where $\m{d}r'$ is the total rest frame distance travelled by the photon.  We use the equations for the synchrotron emission and opacity $k_{\nu}(\nu,x)$ from Paper I with the exception of Equation 23 in Paper I which contains an erroneous extra factor of 2 in the numerator.  Using the formulae for the synchrotron opacity from Paper I we find the observed optical depth to a section to be given by 

\be
\tau_{tot}(\nu,x)=\int^{L}_{x}k_{\nu}(\nu'(x),x) \m{d}r'.
\ee

The observed synchrotron emission from the jet is given by

\be
\nu F_{\nu}=\sum_{x} \delta_{\m{Doppler}}(x)^{4}\nu\rq{} P_{\nu}(x,\m{d}x,\nu') e^{-\tau_{tot}(\nu,x)} ,
\ee

where the length of an emitting section in the rest frame of the plasma $\m{d}x'$ is related by a Lorentz contraction to the lab frame section length $\m{d}x$.

\section{Inverse-Compton emission}

\subsection{SSC emission}

To calculate the SSC emission we use the result of the emission produced in the plasma rest frame from Paper I.  Since the high energy emission is optically thin the only modification to the earlier result is that we need to Doppler-boost the emission from each section individually.  We use Equation 38 from Paper I to calculate the inverse-Compton power emitted by a section in its rest frame per log bin of rest frame frequency (corresponding to $\nu\rq{} F_{\nu}\rq{}$).  

\be
\nu F_{\nu}=\sum_{x} \nu'(x) F_{\nu}\rq{}(\nu\rq{}(x),x,\m{d}x) \delta_{\m{Doppler}}(x)^{4}.
\ee

\subsection{Inverse-Compton scattering of external photons}

In this work we consider the six most likely sources of external photons to the jet: the CMB, radiation directly from the accretion disc, starlight and radiation from the accretion disc scattered by free electrons in the BLR and reprocessed by a dusty torus and NLR.  Inverse-Compton scattering of external photons has been investigated by a number of authors including \cite{1992A&A...256L..27D}, \cite{1993ApJ...416..458D}, \cite{1994ApJ...421..153S}, \cite{2000AJ....119..469B} and \cite{2011MNRAS.415..133H}.  

The full calculation for the inverse-Compton emission of an axisymmetric external photon field by an isotropic electron distribution takes the form of a seven dimensional integral.  In order to make the problem computationally tractable we make a number of simplifying approximations.  Firstly we assume that in a collision with an electron the scattered photon travels with a velocity parallel to the electron's initial velocity.  This approximation is commonly used when considering inverse-Compton scattering of photons (\cite{2009ApJ...692...32D}) and is justified because the scattered photon's energy is a steep function of its scattering angle with significant scattering energies occurring for photons travelling close to the electron's initial velocity, due to conservation of momentum.

For a jet with a significant bulk Lorentz factor radiation which is isotropic relative to the black hole is Doppler-boosted in the rest frame of the jet plasma, such as the CMB, starlight, BLR, NLR and dusty torus photons.  The radiation is Doppler-boosted in the plasma rest frame so that it is beamed along the jet axis in the opposite direction to the jet material with a characteristic opening angle $\sim \frac{1}{\gamma_{\m{bulk}}}$.  This means that for large values of $\gamma_{\m{bulk}}$ almost all external photons travel close to the jet axis.  We shall make the approximation that all the external photons travel in the negative $x$-direction to reduce the dimensionality of the integral to 3.  This approximation is strictly valid only in the limit of large $\gamma_{\m{bulk}}$, however, the scattered energy of the photons depends on the incident photon direction through $\propto (1+\beta \cos(\theta))$ so for a small change in opening angle $\frac{1}{\gamma_{\m{bulk}}}$ the scattered photon energy will not change significantly.  The total number density of external photons is therefore increased by a single Doppler factor and their energy is also increased by a Doppler factor in the rest frame of the plasma (since we have effectively integrated over solid angle).  

In the case of external photons travelling directly from the accretion disc we make similar approximations.  We approximate the scattered photon velocity to be parallel to the initial electron velocity.  We also approximate the external accretion disc photons to travel in the positive $x$-direction, this is justified because the brightest part of the accretion disc is closest to the black hole (at a few Schwarzschild radii) and so the base of the jet is not illuminated isotropically by the accretion disc.  The most energetic photons illuminate the disc from behind in the frame of the black hole. 

Unlike in the case of photons from the CMB, starlight, BLR, NLR and dusty torus, photons directly from the accretion disc are de-boosted, so Lorentz transforming into the plasma rest frame results in the photon angles diverging away from $x$-axis.   This means that our approximation is only justified in the case that the radius of the jet is smaller than the distance from the accretion disc.  If this is not the case then the approximation results in an underestimate of the inverse-Compton emission from accretion disc photons, however, neither the Doppler factor or scattering cross section depend very steeply on the angle of the photons from the $x$-axis for the case of photons illuminating the jet from behind so the results should not differ significantly from a more detailed treatment (see for example \cite{2009ApJ...692...32D}).  Since most previous investigations have assumed de-boosted accretion disc photons are not relevant to inverse-Compton emission it is interesting to see whether this emission is significant.

With these approximations we can use a modified version of the inverse-Compton scattering formula from paper I (Equation 36) to calculate the inverse-Compton emission from external photon sources.  Using these approximations we have eliminated the integration over $\theta$.  Instead we only consider scattering from electrons travelling at an angle $\theta$ where this angle corresponds to photons scattered in the lab frame at the observation angle of the jet $\theta_{\m{obs}}$.  This angle $\theta$ is easily related to the observation angle of the jet via a Lorentz transformation which in the case of CMB, starlight, BLR, NLR or dusty torus photons we calculate to be.

\begin{equation}
\theta=\cos^{-1} \left(\frac{\cos(\theta_{\m{obs}})-\beta(x)}{1-\beta(x) \cos(\theta_{\m{obs}})}\right),
\label{thet1}
\end{equation}

and for the case of external photons directly from the accretion disc

\begin{equation}
\theta=\pi -\cos^{-1} \left(\frac{\cos(\theta_{\m{obs}})-\beta(x)}{1-\beta(x) \cos(\theta_{\m{obs}})}\right).
\label{thet2}
\end{equation}

Fixing the value of $\theta$ using one of Equations $\ref{thet1}$ or $\ref{thet2}$ we find the inverse-Compton emission of external photons to be

\bea
&&\mathrm{weight}(E_{e}, E_{\gamma}, \theta, \phi_{2},x) = N_{e}(E_{e},x).\frac{\rmn{d}\sigma(E_{\gamma}', \phi_{2})}{\rmn{d}\Omega_{2}} \nonumber \\
&& \qquad \, \times n_{\gamma}(E_{\gamma}, \theta, x). c(1-\beta_{e} \cos\theta) \pi |\sin \phi_{2}|.
\eea

\bea
\mathrm{Power}(W) = \sum_{E_{e}, E_{\gamma}, \theta, \phi_{2},x} E_{\gamma \mathrm{scatt}}.\mathrm{weight}(E_{e}, E_{\gamma}, \theta, \phi_{2},x) \nonumber \\
\times \rmn{d}E_{e} \rmn{d}E_{\gamma} \rmn{d}\phi_{2} \rmn{d}x'.
\label{PowIC}
\eea

where $\beta_{e}$ is the electron velocity divided by $c$.  We now proceed to calculate the number density of external photons in the plasma rest frame for the six sources.  In the following subsections quantities defined in the lab frame of the jet are primed and quantities in the rest frame of the jet plasma are not primed.

\subsection{CMB}

We use the blackbody distribution to find $n_{\gamma}$ for the CMB at redshift $z$ with $T=2.73(1+z)\rm{K}$.

\begin{equation}
n'_{\gamma}(\nu')=\frac{4 \pi}{c} \frac{2\nu'^{2}}{c^{2}(e^{\frac{h\nu'}{k_{b}T}}-1)} = \frac{ 8 \pi \nu'^{2}}{c^{3}(e^{\frac{h \nu'}{k_{b}T}}-1)}.
\end{equation}

We average the Doppler factor over solid angle to find the mean Doppler factor for a photon emitted isotropically in the black hole rest frame which is $\gamma_{\m{bulk}}$.  The frequency of the photons and their arrival rate are both boosted by a Doppler factor.  The number density per unit frequency of photons is therefore unaltered since the increase in arrival rate is compensated by the stretching of the frequency bin occupied by the photons.

\be
 \delta_{\m{Dopp}}(x)n_{\gamma}\rq{}(\nu\rq{})d\nu\rq{}=n_{\gamma}(\nu)d\nu, \qquad \delta_{\m{Dopp}}(x)=\gamma_{\m{bulk}},
\ee
\be
n_{\gamma}(\nu)=\delta_{\m{Dopp}}(x) n_{\gamma}\rq{}\left(\frac{\nu}{\delta_{\m{Dopp}}(x)}\right) \frac{d\nu\rq{}}{\delta_{\m{Dopp}}(x)d\nu\rq{}}, \,\, n_{\gamma}(\nu)=n_{\gamma}\rq{}(\nu\rq{}).
\ee

From these equations we find that the total energy density of CMB photons in the plasma rest frame $U_{\m{phot}}=\int h \nu n_{\gamma}(\nu)\m{d}\nu$ is proportional to bulk Lorentz factor squared as we expect from Equation \ref{LTEM}.  

\subsection{Thin Accretion Disc}

We consider a thin accretion disc (\cite{1973A&A....24..337S}) around a black hole of mass $M$ accreting at a rate  $\dot{M}$.  Since we do not know the likely black hole spin parameters of these AGN we will conservatively assume a slowly rotating black hole.  We treat the thin disc as a radiating blackbody.

\be
\rmn{d}L =  \frac{3G \dot{M}M}{2r^{2}} \rmn{d}r = 2\sigma T^{4} 2 \pi r \rmn{d}r ,
\ee
\be
T = \left(\frac {3G\dot{M}M}{8\sigma \pi r^{3}}\right)^{\frac{1}{4}} ,
\ee
\be
L_{\nu} = \int_{r_{in}}^{r_{out}} \frac{4\pi h r \nu'^{3}}{c^{2}\left(e^{\frac{h\nu'}{k_{b}T}-1}\right)} \rmn{d}r, \qquad r_{in} = \frac{6GM}{c^{2}}.
\ee

We make the approximation that the accretion disc luminosity is emitted isotropically and consider the photon density at a radius $R$ from the disc in the lab frame. 

\begin{equation}
n'_{\gamma}=\frac{L_{\nu'}}{4 \pi R^{2}h\nu' c}.
\end{equation}

The photon number density and frequency both decrease by a Doppler factor when they are Lorentz transformed into the plasma rest frame.  Again we use the value of the Doppler factor averaged over solid angle

\be
n_{\gamma}(\nu)=n'_{\gamma}(\nu\delta_{\m{Dopp}}(x)), \,\,\,\,\, \delta_{\m{Dopp}}(x)=\gamma_{\m{bulk}}.
\ee

\subsection{Starlight}

Inverse-Compton scattering of starlight by jets has been investigated previously by \cite{2011MNRAS.415..133H}.  In this work we are fitting to a powerful Compton-dominant blazar so we expect its host galaxy to be a giant elliptical.  Work by \cite{1995AJ....110.2622L} and \cite{1997AJ....114.1771F} found that the luminosity density at the centre of giant ellipticals is considerably lower than that of smaller elliptical or disk galaxies with the luminosity density at 30pc in the most massive ellipticals found to be typically between $2-10 L_{\odot}\m{pc}^{-3}$.  These large ellipticals have relatively flat cores over which the luminosity density goes as $\rho_{L}\propto r^{a}$ with $-1.3<a<-1.0$ (\cite{1995AJ....110.2622L}), at larger distances the luminosity density decreases as $\rho_{L}\propto r^{-4}$ (\cite{1994AJ....107..634T}).  In M87 the flat core with a luminosity density power law index $a=-1.3$ extends out to ~40kpc (\cite{1997ApJ...486..230C}).  Using the typical values above to model the seed photon population due to starlight along the jet we choose a luminosity density given by

\be
\rho_{L}(r)=\frac{5L_{\odot} (30\pc)^{1.2}}{r^{1.2}}(\W\pc^{-3}) \,\,\,\, \m{for}\,\,\,\, 1\pc\geq r\leq40\kpc,
\ee
\be
\rho_{L}(r)=\frac{\rho_{L}(r=40\kpc) (40\kpc)^{4}}{r^{4}}(\W\pc^{-3}) \,\,\,\, \m{for} \,\,\,\,  r>40\kpc,
\ee

where the luminosity density is assumed to be constant within the central 1pc of the galaxy.  We calculate the total enclosed stellar mass in the core of the galaxy corresponding to this luminosity density to be $4.0\times 10^{12}M_{\odot}$ using a mass to light ratio of 10, this is consistent with the expected stellar mass in the core of a large elliptical galaxy.  We calculate the energy density of starlight observed by the jet plasma at a distance $x$ along the jet ($\rho_{\m{star}}(x)$) by the integral

\be
\rho_{\m{star}}(x)=\int_{x}^{100\m{kpc}}\frac{4 \pi r^{2} \rho_{L}(r)}{4\pi r^{2} c}\m{d}r=\int_{x}^{100\m{kpc}} \frac{\rho_{L}(r)}{c} \m{d}r,
\ee

where we have started the integral at the position of the jet section, x, since the starlight from regions $r<x$ will illuminate the jet from behind and so will be strongly Doppler-deboosted.  We have also made use of the spherical symmetry of the problem to simplify the integral, since all observers inside a spherically symmetric emitting surface see the same isotropic luminosity and surface brightness (due to Newton\rq{}s shell theorem) we are free to calculate the observed energy density of an observer at the origin to simplify the integral since this is the same as the luminosity observed by the plasma at $r=x$.  Using this integral we find the energy density of starlight at a distance of 1pc from the centre of the galaxy to be $\approx 2\times10^{-12}\m{Jm}^{-3}$.  This is in agreement with the value quoted in \cite{2011MNRAS.415..133H} for the energy density in starlight at the centre of an elliptical galaxy.

We assume that the starlight is predominantly due to K-type stars which typically dominate the spectra of elliptical galaxies.  We choose the photon distribution to be that of a 5000K blackbody since this corresponds to the surface temperature of a typical K-type star.  In practice it is difficult to accurately model the interstellar radiation field at the centre of a powerful FSRQ since the detailed properties of the stellar population will vary substantially between galaxies.  We include starlight in our model for completeness and to investigate whether there is evidence of inverse-Compton scattering of starlight from fitting to the spectrum of PKS0227.

\subsection{Accretion disc photons scattered by the broad line region}

To calculate the number density of accretion disc photons scattered by the broad line region in the lab frame we use the result from \cite{2009ApJ...692...32D}.  

\be
n'_{\gamma}(E'_{\gamma}) \approx \frac{n_{e}^{\m{BLR}}(R)\sigma_{\m{T}}L_{\m{acc}}(E'_{\gamma})}{4\pi RcE'_{\gamma}},
\ee

where $n_{\m{e}}^{\m{BLR}}(R)$ is the free electron number density.  The authors calculated this result assuming that accretion disc photons are Thomson scattered by free electrons in the BLR.  The result assumes that the scattered radiation field is isotropic in the lab frame so the photon number density and energy in the plasma rest frame are related to that in the lab frame by a Lorentz transformation.  

\be
n_{\gamma}(E_{\gamma})= n'_{\gamma}\left(\frac{E'_{\gamma}}{\delta_{\m{Dopp}}(x)}\right), \,\,\,\,\, \delta_{\m{Dopp}}(x)=\gamma_{\m{bulk}}. \label{ngamrest}
\ee

where we have used the value of the Doppler factor averaged over solid angle.  We assume the BLR extends out to $R_{\m{BLR}}=0.019 \left(\frac{\lambda L_{\lambda}(5100 \AA)}{10^{44}\m{erg}\,\m{s}^{-1}}\right)^{0.69}$pc (\cite{2005ApJ...629...61K}), which we take to be 0.5pc for the most powerful FSRQ type blazars such as PKS0227.  To calculate the value of $n_{e}^{\m{BLR}}$ we assume a constant number density of electrons out to $R_{\m{BLR}}$ and a covering fraction for the BLR, $\eta_{\m{BLR}}=0.1$ (\cite{MNR:MNR13360} and \cite{2009ApJ...704...38S}).  The covering fraction is given approximately by

\be
\eta_{\m{BLR}}=\int \frac{ n_{e}^{\m{BLR}}\sigma_{\m{T}}}{4 \pi R^{2}}4 \pi R^{2}\m{d}R \approx n_{e}^{\m{BLR}}\sigma_{\m{T}}R_{\m{BLR}}.
\ee

For our chosen covering fraction and BLR radius this gives $n_{e}^{\m{BLR}}= 1.0 \times 10^{11}\m{m}^{-3}$.

\subsection{Dusty Torus and Narrow Line Region}

We now calculate the number density of photons scattered by clumps of dust in the torus and narrow line region (NLR) surrounding the black hole.  At a distance $x$ from the black hole the surface temperature of the dust facing towards the accretion disc, $T_{\m{dust}}$, is given by (\cite{2008ApJ...685..147N})

\be
T_{\m{dust}}=1500\left(\frac{x}{x_{d}}\right)^{0.39}K  \qquad x \leq 9 x_{d},\nonumber
\ee
\be
T_{\m{dust}}=640\left(\frac{9x}{x_{d}}\right)^{0.45}K  \qquad x>9x_{d}. \label{Tdust}
\ee

where $x_{d}=0.4(L_{\m{acc}}/10^{37}W)^{1/2}$pc is the dust sublimation distance and $L_{\m{acc}}$ is the total accretion disc luminosity.  We assume a clumpy dust cloud distribution where the number density of clouds is given by $n_{\m{cloud}}\propto x^{-q}$ with $1\leq q\leq 2$ following \cite{2008ApJ...685..147N}.  We also assume that the distribution of cloud sizes is independent of $x$.  The hot torus absorbs and reprocesses a fraction (the covering fraction) $\eta_{T}$ of the incident accretion disc radiation.  We use a covering fraction of $\eta_{T}=0.1$ taken from the ratio of accretion disc to dusty torus emission in \cite{2008ApJ...685..160N}.  We further assume that this hot torus extends from the dust sublimation distance $x_{d}$ out to $\sim 10x_{d}$ as indicated by observations (\cite{2006NewAR..50..728E} and \cite{2008ApJ...685..160N}).  We assume that the NLR has a covering fraction $\eta_{\m{NLR}}=0.07$ (\cite{2009ApJ...705..298M}) and extends from the outer radius of the torus ($x=10x_{d}$) out to a distance $x_{\m{outer}}$, typically around 1kpc (\cite{2006A&A...459...55B}).  

We now calculate the number density of photons from this dust distribution.  Taking an average cross-sectional area of a cloud $\sigma_{cloud}$ independent of $x$, the cloud number density and covering fraction we can calculate the emissivity of the dust as a function of distance from the black hole.  The average emissivity of the dust $j(x)$ at a distance $x$ will be proportional to the number density of clouds at $x$, so $j(x)=j_{0}x^{-q}$

\bea
\eta_{T}L_{\m{acc}}=\m{const.} \int_{x_{d}} ^{10x_{d}}&& n_{\m{cloud}}(x) \sigma_{\m{cloud}} 4\pi x^{2} \m{d}x= \nonumber \\
&& j_{0} \int_{x_{d}} ^{10x_{d}} 4\pi x^{2} x^{-q} \m{d}x.
\eea 

From this equation we can calculate $j_{0}$ for both the torus and NLR (with the appropriate torus and NLR quantities interchanged).  Since the bulk Lorentz factor of the jet is large we only consider emission from spherical shells at a radius greater than the distance of the jet section from the black hole, since radiation emitted from behind the jet is Doppler-deboosted in the rest frame of the jet.  We use Newton's shell theorem to calculate the energy density of radiation inside an emitting shell of dust.  We assume clouds re-emit the absorbed accretion disc radiation as a black body with a temperature given by Equation \ref{Tdust} (this is the correct dust temperature since we are only interested in the emission from the surface of clouds facing back towards the jet and black hole).  In this case the number density of photons from dust outside a distance $x$ in the lab frame is given by

\be
n'_{\nu}(x,\nu')=\int_{x} ^{x_{\m{outer}}} \frac{j(\nu',x)}{4\pi x^{2}ch\nu'} 4 \pi x^{2}\m{d}x,
\ee

where $j(x, \nu')$ is the appropriate blackbody distribution normalised to the correct emissivity.

\be
j(x,\nu')=\frac{\pi j_{0}x^{-q}}{\sigma T_{\m{dust}}(x)^{4}}. \frac{2h \nu'^{3}}{c^{2}}. \frac{1}{e^{h\nu'/k_{B}T_{\m{dust}}(x)}-1}.
\ee  

The number density of photons in the plasma rest frame is then given by

\be
n_{\nu} (x,\nu)=n'\left(x,\frac{\nu}{\delta_{\m{Dopp}}}\right),\qquad \delta_{\m{Dopp}}=\gamma_{\m{bulk}}(x).
\ee

In this work we take the value of $q=1$ for the cloud distribution as suggested by \cite{2009ApJ...705..298M} from fitting to observations.

\subsection{Observed external Compton emission}

The observed inverse-Compton emission from the external photons is related to the emission in the rest frame by

\be
\nu F_{\nu}=\sum_{x} \nu'(x) F_{\nu}\rq{}(\nu\rq{}(x),x,\m{d}x) \delta_{\m{Doppler}}(x)^{4}.
\ee

\section{Particle acceleration and energy losses}

We take into account radiative and adiabatic losses to the electron population as it traverses along the jet.  We use Equation $\ref{loss}$ to take into account the effects of radiative losses to the population due to both synchrotron and inverse-Compton emission.  In addition to radiative losses in this investigation we will also consider adiabatic losses in the accelerating parabolic section of the jet.  We assume the conical section of the jet is ballistic and so does not suffer adiabatic losses as in Paper I.  We use the formula from \cite{2006MNRAS.367.1083K} to include adiabatic losses to a relativistic electron population.   

\be
E'=E\left(\frac{V'}{V}\right)^{-\frac{1}{3}}=E \left(\frac{R}{R'}\right)^{\frac{2}{3}}, \label{Ead}
\ee
\be
N'(E')=N(E),
\label{ad}
\ee

where E is the initial energy of an electron, V is the initial volume of the section and R the initial radius, primed quantities are those after expansion.  Considering the electron distribution contained in a slab of width one metre in the rest frame we use the following equation to take into account radiative energy losses to the electron population as it travels along the jet.

\be
N_{e}(E_{e},x+\gamma_{\m{bulk}}\m{d}x')=N_{e}(E_{e},x)-\frac{P_{tot}(x,\m{d}x',E_{e}) \times 1s}{c E_{e}}, \label{loss}
\ee

where $P_{tot}$ is the total power emitted by electrons of energy $E_{e}$ in a section of rest frame jet width $\rm{d}x'$ due to synchrotron and inverse-Compton emission.  The energy losses for the synchrotron and SSC emission are the same as those calculated in Paper I.  We have simplified the calculation of the inverse-Compton scattering of external photons by considering only the scattering events which lead to photons travelling at the observation angle.  To calculate the electron energy losses from this anisotropic photon field we must take into account the average energy lost from isotropic electrons encountering a narrow beam of external radiation.  In this case we can use the formula in Paper I for the total power emitted by an isotropic electron distribution scattering an isotropic photon field using the value of $n_{\gamma}$ calculated above for the external photon fields.  After radiative losses have been taken into account we then apply Equation \ref{ad} to the population to take into account adiabatic losses in the accelerating parabolic section.  

\section{Acceleration of electrons in the decelerating conical part of the jet}

Observations of AGN jets show evidence for in situ acceleration occurring along the jet.  High energy optical synchrotron electrons are observed at distances beyond that expected from their synchrotron lifetime (see \cite{1997A&A...325...57M} and \cite{2001A&A...373..447J}).  Observations also indicate that jet material decelerates away from the base though the jet remains relativistic until the hot spot (\cite{2002MNRAS.336..328L}, \cite{2005MNRAS.358..843H} and \cite{2010ApJ...710..743D}).  In our model shocks in the jet convert bulk kinetic energy into electron energy in the rest frame of the plasma as the jet decelerates once it has become conical.  This is a natural result of conservation of energy-momentum in our model since a decelerating jet loses kinetic energy and so must gain internal energy (or do work on its environment) to conserve energy.  We have assumed that the excess bulk energy is converted into both particle acceleration and amplification of the magnetic field to maintain equipartition throughout the conical section of the jet.  

In this work we simulate the effect of acceleration of electrons by the injection of a fixed power law distribution of electrons continuously along the jet which conserves energy-momentum in the decelerating part of the jet.  Observations of AGN jets show a continuous variation in optical spectral index along the jet with evidence for continuous acceleration taking place along the jet (\cite{2001A&A...373..447J}) and so converting bulk kinetic energy to internal electron energy through deceleration is a natural way in which to explain these observations.  The acceleration of these electrons is thought to be caused by shock acceleration where the shocks originate from a variety of mechanisms including the collision of shells of plasma travelling at different speeds as the jet decelerates (\cite{2001MNRAS.325.1559S}), shear acceleration (\cite{2002A&A...396..833R}) or external pressure driven recollimation shocks (\cite{1998MNRAS.297.1087K}).  Variations of models in which in situ acceleration of electrons occurs are popular and have been investigated by many authors including \cite{2000A&A...356..975K}, \cite{2001MNRAS.325.1559S} and \cite{2010MNRAS.401..394J}.  This is the first time, however, that conversion of bulk kinetic energy into accelerating electrons through deceleration has been considered self-consistently in a continuous conical jet with a variable Lorentz factor. 

We assume that the injected electrons have the same spectrum as the initial electron population (Equation $\ref{ne}$).  In this investigation we will assume that the steady-state shock distribution is the same for all shocks along the jet.  In practice the power law exponent and maximum energy cutoff will depend on the detailed properties of the shock and plasma (\cite{1978MNRAS.182..147B}) but this is beyond the scope of our current investigation.

We calculate the evolution of the electron distribution along the jet by taking into account radiative energy losses on the electron population (and adiabatic losses on the remaining electrons in the parabolic section).  We determine the energy density of electrons injected into the plasma due to deceleration by Equation $\ref{cont1}$.

\section{Equipartition considerations}

Observations indicate that on large scales the jet plasma is close to equipartition (\cite{2005ApJ...626..733C}).  Observations of brightness temperatures of the cores of jets also indicate that the jet plasma is close to equipartition when sources are in a quiescent state, however, in some cases shocks are believed to produce regions where the plasma becomes particle dominated associated with flaring (\cite{2006ApJ...642L.115H}).  It is thought that jets are produced via the interaction of electromagnetic fields with a rotating black hole (\cite{1977MNRAS.179..433B}).  Current models of jet production start with a magnetically dominated region which accelerates particles to form the jets we observe.  Simulations indicate that the jet becomes conical when the plasma nears equipartition (\cite{2006MNRAS.368.1561M} and \cite{2009MNRAS.394.1182K}).  This makes physical sense since the magnetic field has done work accelerating the electron plasma and in equipartition the remaining magnetic energy is not large enough to accelerate the jet substantially.  In this model we force the jet plasma to be in equipartition after the point at which the jet becomes conical by the injection of shock accelerated electrons.  

For a given total jet power we expect the synchrotron and SSC emission to be brightest when the jet is close to equipartition from a reversal of the minimum energy argument for synchrotron emitting lobes (\cite{1959ApJ...129..849B}).  If the accelerating region is magnetically dominated we therefore expect the synchrotron emission to be brightest at the point where the jet becomes conical since the bulk Lorentz factor, jet power and magnetic field will be largest at this point and decrease along the conical section.  

\section{Code}

We use the equations given in the preceding sections and Paper I to calculate the emission from the electrons along the jet.  In order to make the problem computationally tractable we must discretise the electron population, sections of the jet and the particle acceleration resulting from the deceleration.  Our code uses logarithmic bins of electron energy where the bin of energy $E_{e}$ has a width $\m{d}E_{e}=0.3E_{e}$, so the total electron population for PKS0227 spans approximately 40 energy bins and for high frequency peaked BL Lacs this increases to approximately 60 bins.  The jet sections are determined adaptively so that either the shortest radiative lifetime is resolved or the jet radius does not increase by more than $5\%$, this results in jets typically being composed from 200-600 cylindrical sections.  At each decelerating section the equipartition fraction of the jet is calculated and additional electrons ($\Delta \rho_{\m{dec}}$ in Equation \ref{cont1}) are only injected if this equipartition fraction differs from 1 by more than $10\%$ so that we are not continually limited to resolving the shortest electron radiative lifetime.  

We have chosen the above values for discretisation so that increasing the resolution/number of bins does not noticeably affect the spectrum produced from the jet, whilst this allows our code to produce a spectrum for a jet in 15-30 seconds running on a single processor core.  

\section{Results}

\begin{figure*}
	\centering
		\includegraphics[width=15 cm, clip=true, trim=1cm 1cm 0cm 1cm]{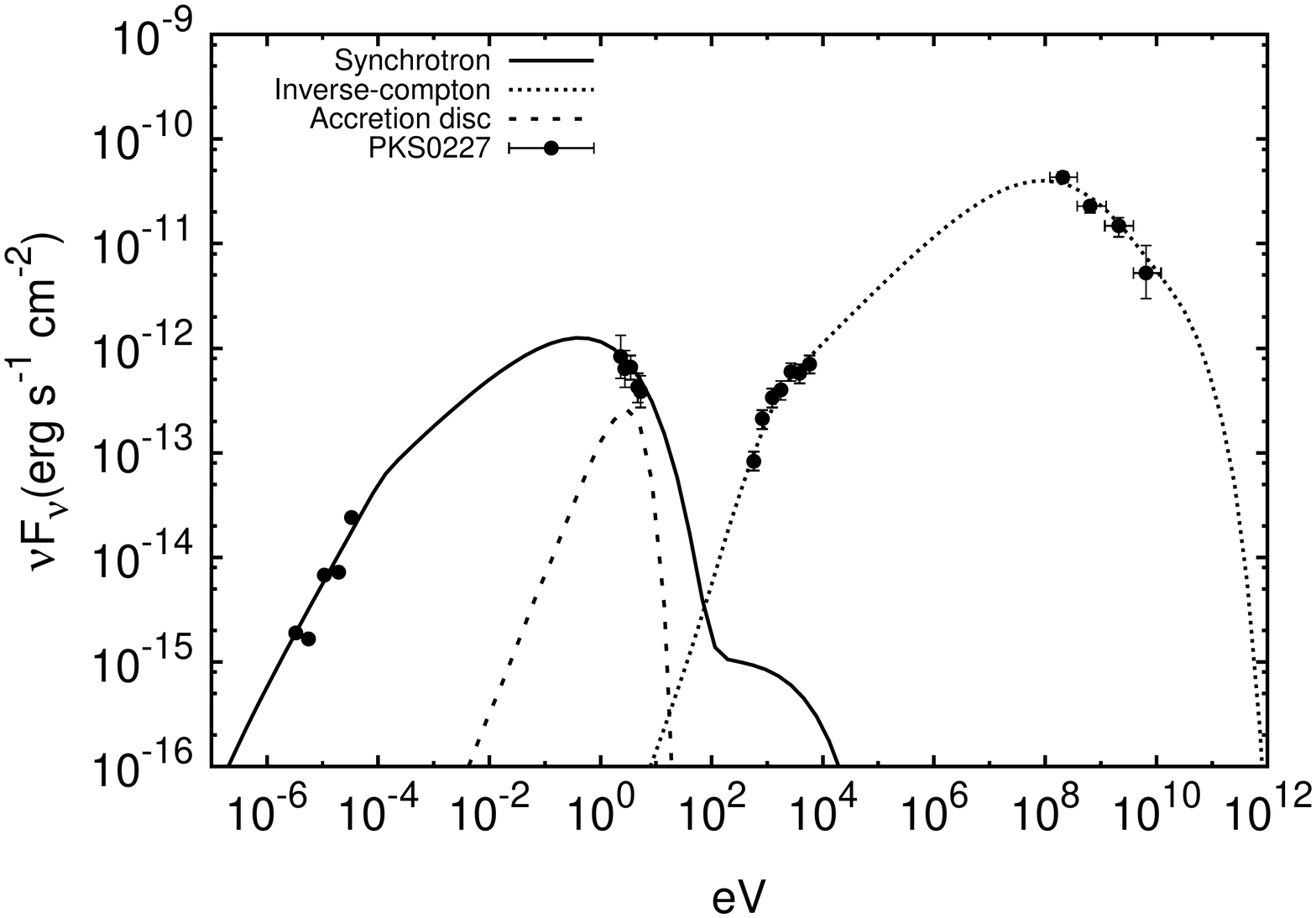}
			
	\caption{This figure shows the results of fitting the model to the SED of PKS0227. The model fits the observations very well across all wavelengths including radio observations.  We find the model fit requires a high jet power, high Lorentz factor jet observed at a small angle to the jet axis.  We find that the inverse-Compton peak is due to scattering of Doppler-boosted high-redshift CMB with a small contribution from scattering NLR photons at the highest gamma-ray energies. }
	\label{fig1}
\end{figure*}
\begin{figure*}
	\centering
		\subfloat[The fit to PKS0227 showing the different contributions to the inverse-Compton emission.]{ \includegraphics[width=8 cm, clip=true, trim=1cm 1cm 0cm 1cm]{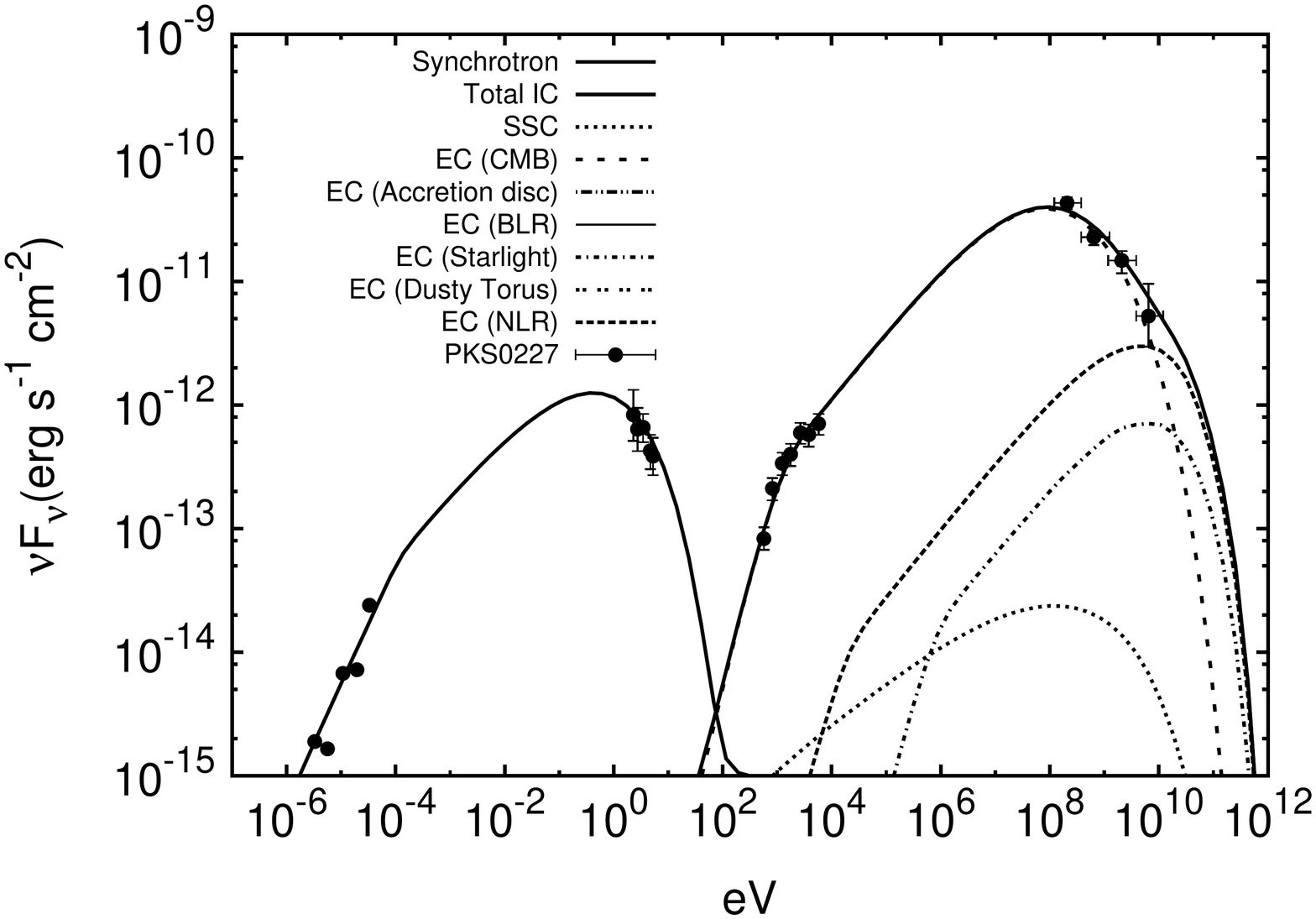} } 
		\qquad
		\subfloat[The fit to PKS0227 showing the emission from different distances along the jet]{ \includegraphics[width=8 cm, clip=true, trim=1cm 1cm 0cm 1cm]{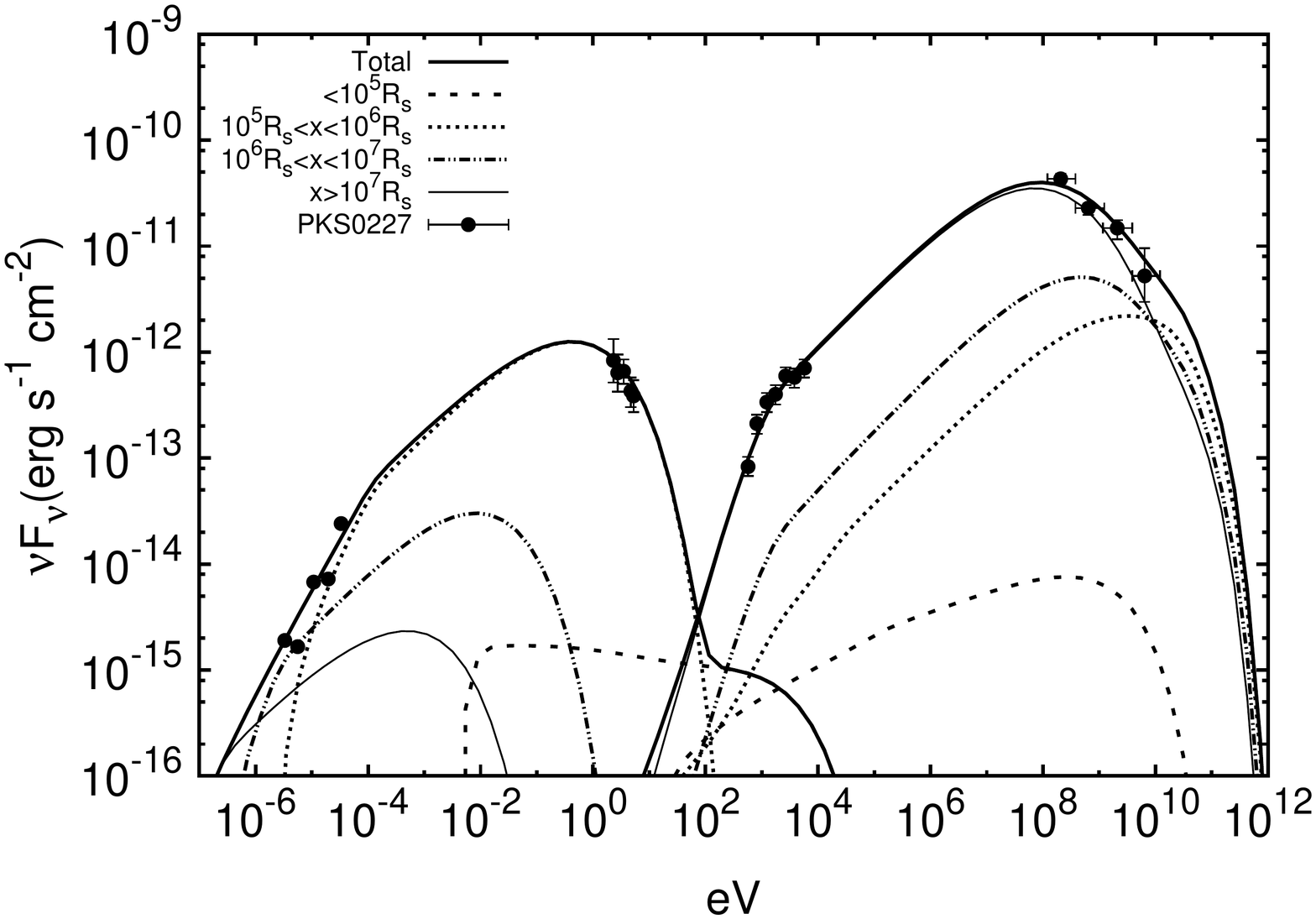} }
	
	\caption{This figure shows the different inverse-Compton components in Figure a and the jet emission from different distances in Figure b, using a transition region at large distances.  Figure a shows that the inverse-Compton emission is produced by scattering CMB photons with a small contribution from NLR at the highest gamma-ray energies.  Figure b shows that the optically thin synchrotron emission is dominated by the transition region, with the radio emission coming from larger distances as we expect.  The inverse-Compton emission originates at large distances from the central black hole from scattering CMB photons.}
	\label{fig2}
\end{figure*}

In this paper we have set out to develop a physically realistic jet model which is consistent with observations and simulations and includes all the relevant physics to accurately predict its emission.  In order to test whether the model can potentially describe real blazars we shall try to fit to the spectrum of the most Compton-dominant FSRQ from (\cite{2010ApJ...716...30A}) PKS0227-369.  We choose a Compton-dominant blazar since in Paper I we showed that a simpler version of this model was able to reproduce the spectrum of BL Lacertae, a relatively weak powered object.  Compton-dominant blazars are thought to represent high-powered blazars (\cite{1998MNRAS.299..433F}) so it is interesting to see if our model is capable of reproducing the spectra of the two extremes of the blazar population.  We assume a standard $\Lambda$CDM cosmology (\cite{2011ApJS..192...18K}) to calculate the distance to PKS0227 which has a redshift of $z=2.115$ (\cite{2010ApJ...716...30A}).

We will first show the results of fitting our model to the data using a transition region at different distances along the jet at large distances $>10$pc, within the dusty torus $\sim$pc and within the BLR $<0.5$pc.  One of the most important current questions about jets is where the gamma-ray emission originates from.  Using our realistic continuous jet model we attempt to fit the quiescent jet emission of PKS0227 and try to constrain the location of the quiescent jet emission.  

\subsection{Fit 1 - A transition region at large distances}

\begin{figure*}
	\centering
		\includegraphics[width=15 cm, clip=true, trim=1cm 1cm 0cm 1cm]{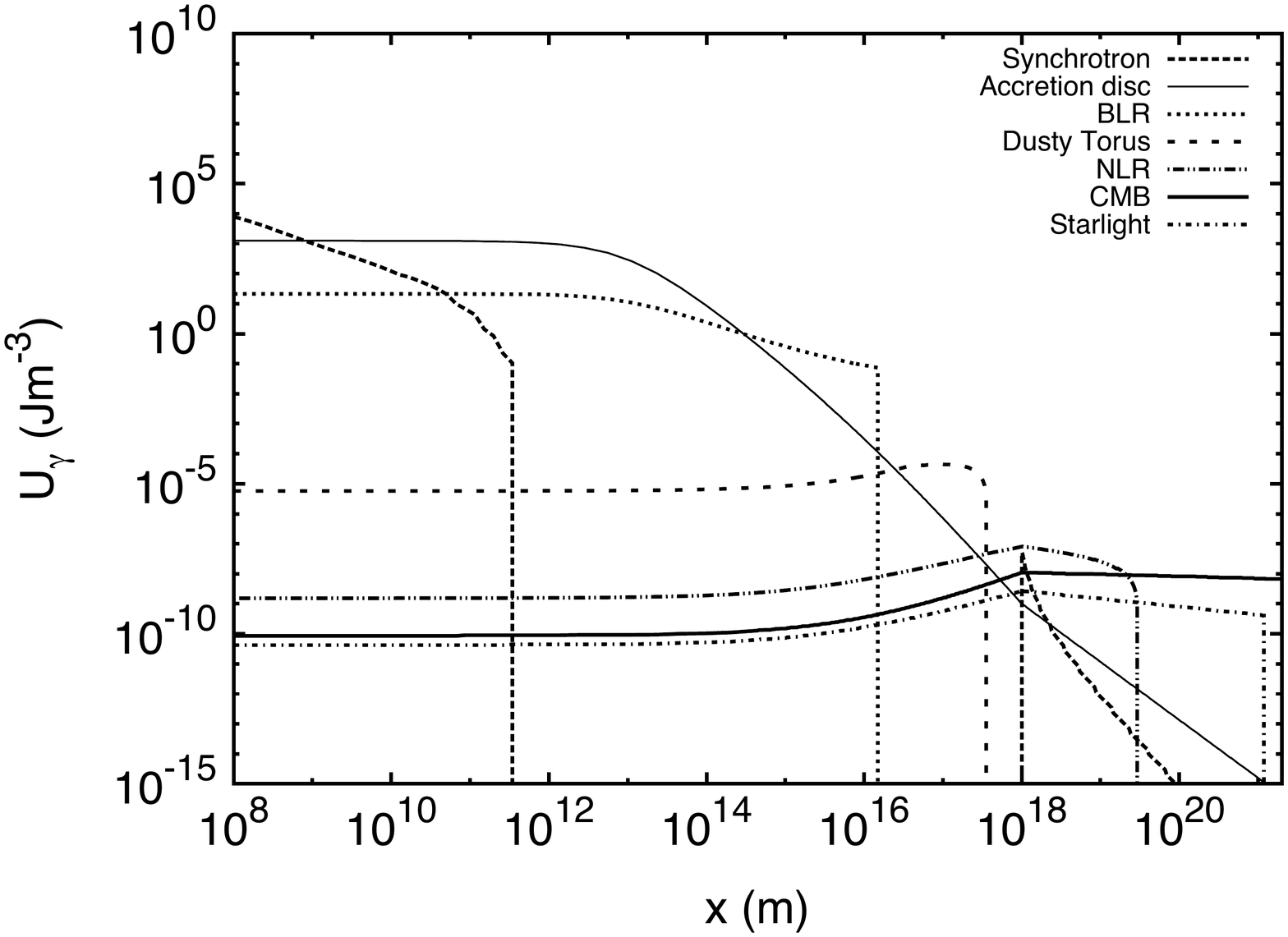}
			
	\caption{This figure shows the energy density of the different radiation sources in the rest frame of the jet at different distances along the jet for Fit 1 to PKS0227.  We see that the acceleration of the jet results in a steady increase in the Doppler-boosting of the BLR, dusty torus, NLR, CMB and starlight up to the transition region.  At different distances along the jet different external photon sources become dominant as we expect.  At the transition region ($x=1.02 \times 10^{18}$m) the NLR is the dominant energy density resulting in the highest energy gamma-ray emission, however, the CMB becomes the dominant energy density at the largest distances along the jet where the majority of inverse-Compton emission is produced. }
	\label{fig3}
\end{figure*}

\begin{table*}
\centering
\begin{tabular}{| c | c | c | c |}
\hline
Parameter & Fit 1 & Fit 2 & Fit 3 \\ \hline
$W_{j}$ & $1.2 \times 10^{39}\m{W}$ & $1.5 \times 10^{39}\m{W}$ & $1.2 \times 10^{39}\m{W}$ \\ \hline
$L_{\m{acc}}$ & $8.3 \times 10^{38}W$ & $8.3 \times 10^{38}W$ & $8.3 \times 10^{38}W$ \\ \hline
$M_{\m{acc}}$ & $3.4 \times 10^{9}M_{\odot}$ & $1.5 \times 10^{8}M_{\odot}$ &  $2.0 \times 10^{9}M_{\odot}$ \\ \hline
L & $2 \times 10^{21}m$ & $ 10^{21}m$ & $ 2 \times 10^{21}m$ \\ \hline
$\theta_{\m{obs}}$ & $0.8^{o}$ & $0.8^{o}$ & $1.5^{o}$ \\ \hline
$\theta_{\m{opening}}$ & $3^{o}$ & $3^{o}$ & $3^{o}$  \\ \hline
$A_{\m{equi}}(x=0)$ & 0.01 & 0.01 & 0.01 \\ \hline
$M$ & $3.4 \times 10^{9}M_{\odot}$ & $1.5 \times 10^{8}M_{\odot}$ &  $4.5 \times 10^{7}M_{\odot}$\\ \hline
$\alpha$ & 1.9 & 1.7 & 1.5 \\ \hline
$E_{\m{min}}$ & $10.2 \,\m{MeV}$ & $15.3 \,\m{MeV}$ & $102 \,\m{MeV}$\\ \hline
$E_{\m{max}}$ & $7.22  \,\m{GeV}$ & $643  \,\m{MeV}$ & $435  \,\m{MeV}$\\ \hline
$\gamma_{0}$ & 4 & 4 & 4 \\ \hline
$\gamma_{\m{max}}$ & 45 & 48 & 15 \\ \hline
$\gamma_{\m{min}}$ & 35 & 38 & 10 \\ \hline
$x_{\m{outer}}$ & $1 \m{kpc}$ & $1 \m{kpc}$ & $1 \m{kpc}$ \\ \hline
\end{tabular}
\caption{This table shows the values of the physical parameters used in the model fits to the multi-wavelength observations of PKS0227 as shown in Figures $\ref{fig1}$ and $\ref{fig4}$ and \ref{fig6}.  }
\label{tab1}
\end{table*}

We first attempt to fit the spectrum using a transition region at large distances outside of the dusty torus and BLR as implied by the geometry of M87 where the jet transitions from parabolic to conical at $10^{5}R_{s}$, corresponding to a distance of $66$pc.  

In Figure $\ref{fig1}$ we show the results of fitting the model by eye to the simultaneous multi-wavelength spectrum of PKS0227.  We see that the model fits very well to the observations across all wavelengths including radio data points with a transition region at a lab frame distance of $34$pc and a jet radius of $0.64$pc.  In Table $\ref{tab1}$ we show the model parameters for this fit to PKS0227.  We find that the fit to the observations requires a high power ($W_{j}=1.2 \times10^{39}W$), high bulk Lorentz factor jet ($\gamma_{\m{max}}=45$), observed at a small angle to the line of sight ($\theta_{\m{obs}}=0.8^{o}$).  This is consistent with our expectations of Compton-dominant FSRQs being FRII type AGN observed close to the jet axis (\cite{1995PASP..107..803U}).  We find that the jet has a maximum bulk Lorentz factor of $45$ and the conical section of the jet decelerates to a bulk Lorentz factor of $35$ by $x=2 \times 10^{21}$m.  For these parameters the magnetic field strength in the plasma rest frame at the transition region is relatively low, $2.03\times10^{-6}T$, required by the low synchrotron peak frequency and Compton-dominance of the blazar.  We find a value of the electron spectral index $\alpha=1.9$ close to the value expected from standard diffusive shock acceleration. 

At these large distances the Compton-dominance is due to inverse-Compton scattering of CMB seed photons with a small contribution by NLR photons at the highest gamma-ray energies (see Figure \ref{fig2}a).  In Figure \ref{fig3} we show the energy density of the different radiation fields measured in the plasma rest frame at different distances along the jet for this fit.  We see that the NLR dominates at the transition region but the CMB is the dominant photon source at large distances along the jet.

We include starlight seed photons in our model, however, we find that inverse-Compton emission from the CMB dominates the emission from starlight, as shown in Figure $\ref{fig2}\m{a}$, except at the highest energies.  This is because the energy density in the CMB increases as $\propto(1+z)^{4}$, so whilst the energy density in starlight is significantly larger at $z=0$, we find that the CMB photons are more significant for high-redshift Compton-dominant blazars such as PKS0227 (see Figure \ref{fig3}).  The inverse-Compton scattering of starlight is also reduced relative to that of the CMB due to the decreased Klein-Nishina cross section for the higher energy starlight photons.  

We find that accretion disc photons directly from the disc are the dominant photon energy density only close to the base of the jet (within a few $R_{S}$ of the black hole) and quickly become dominated by other photon sources due to the Doppler-deboosting and the $~\frac{1}{x^{2}}$ dependence of the photon number density (see Figure \ref{fig3}).  We find that for our fit to PKS0227 the Compton-dominance is due to inverse-Compton scattering of CMB and NLR photons in the conical jet region, as shown in Figure $\ref{fig2}\m{a}$.  Previous investigations tend to ignore the contribution of CMB photons to the inverse-Compton emission, this is because in a BL Lac type object with a low bulk Lorentz factor jet at low redshift the CMB is dominated by SSC photons.  We find that due to the high bulk Lorentz factor and high redshift of our model fit the CMB becomes the dominant photon source at large distances along the jet, $>x_{\m{outer}}$, when taking into account the Doppler-boosting of the CMB into the rest frame of the plasma and including the redshift dependence of the CMB number density.

This is an interesting result since it has implications for the Cosmic evolution of Compton-dominant blazars.  If their Compton-dominance is due to inverse-Compton scattering of CMB photons then we might expect to see a redshift dependence of the Compton-dominance of blazars, with increasing Compton-dominance at higher redshift (since the energy in the CMB scales as $(1+z)^{4}$).  This could explain why most Compton-dominant blazars are observed at high redshifts.  We might also expect to see rarer high powered, high Lorentz factor blazars at large redshift due to the increase in comoving volume and due to the larger number of major mergers in elliptical galaxies which host FRII sources.  In practice it may be difficult to disentangle all these evolutionary effects, however, we might expect that high Lorentz factor blazars at low redshift are less Compton-dominant than the same sources at high redshift where the CMB has a higher blackbody temperature.  

Figure \ref{fig2}b shows the total emission from different sections of the jet.  The synchrotron emission is dominated by the transition region of the jet due to its high bulk Lorentz factor and higher magnetic field strength than outer parts of the jet.  The low frequency radio emission originates from farther along the jet where the jet radius is larger.  The inverse-Compton emission is dominated by the transition region and outer parts of the jet where electrons are accelerated through shocks in the decelerating part of the jet and radiate this energy slowly, primarily through scattering NLR photons at distances $x<x_{\m{outer}}$ and CMB photons at larger distances. 

\begin{figure*}
	\centering
		\includegraphics[width=15 cm, clip=true, trim=1cm 1cm 0cm 1cm]{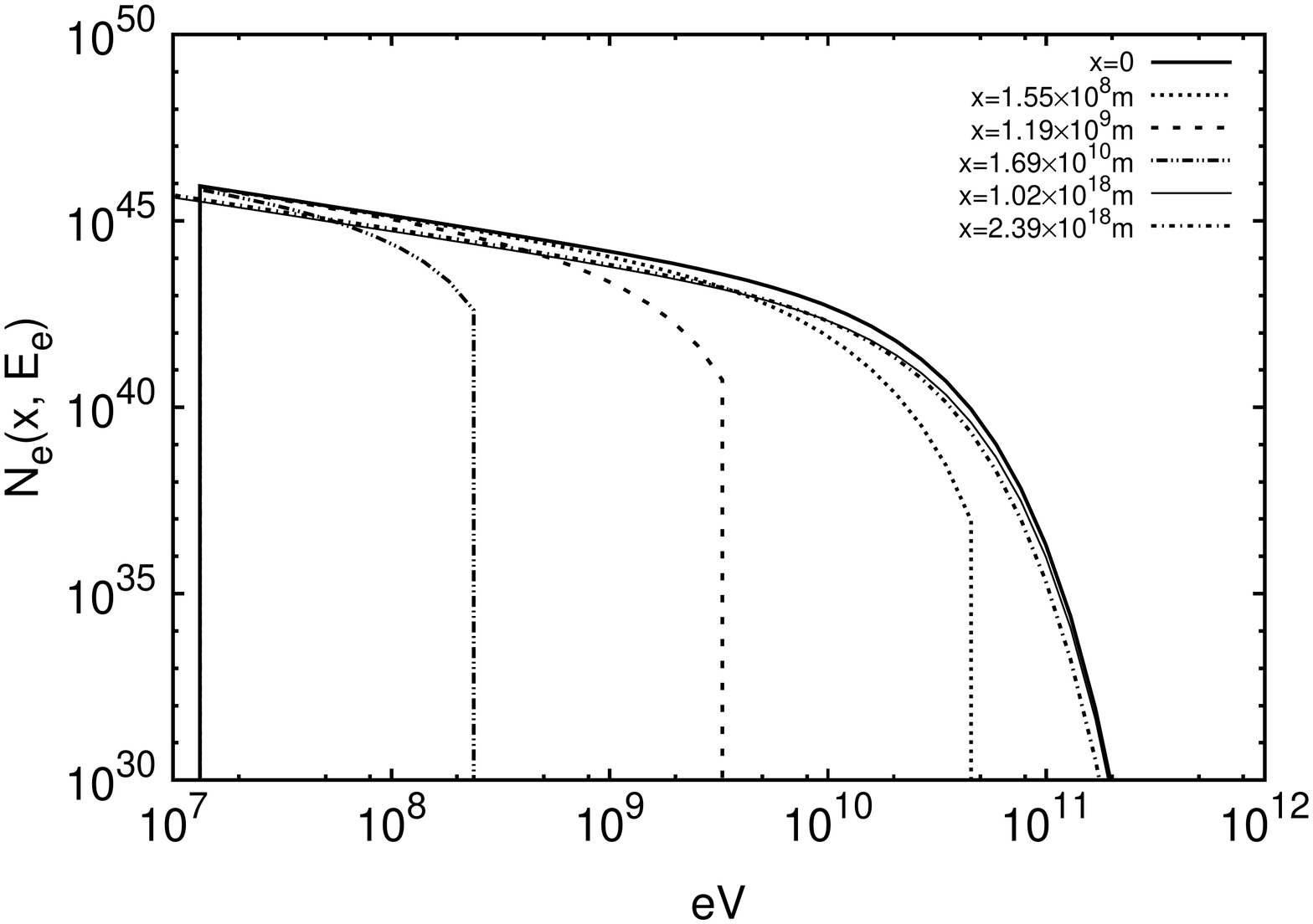}
			
	\caption{This figure shows the evolution of the electron population contained in a slab of width 1m in the plasma rest frame as it travels along the jet in Fit 1.  At the base of the jet, at $x=0$, the electron population given by Equation \ref{ne}, the electrons suffer large radiative and adiabatic losses in the compact accelerating region due to the high magnetic field strength and so the population quickly cools with the highest energy electrons cooling fastest.  The population cools until it reaches the shock at the base of the conical section at $10^{5}R_{s}=1.02\times10^{18}$m where additional electrons are accelerated and the plasma is in equipartition.  The population then proceeds to cool radiatively as it moves away from the shock but with a substantially greater radiative lifetime than at the base of the jet.  This behaviour is repeated as the electrons are accelerated in the decelerating part of the jet.}
	\label{fig9}
\end{figure*}

In Figure \ref{fig9} we show the evolution of the electron population in a slab as it travels along the jet in fit 1.  A small amount of energy is injected into electrons at the base of the jet which quickly cool both radiatively and adiabatically from high to low energies due to the large magnetic field strength.  The population is replenished by the shock at the base of the conical section located at $10^{5}R_{s}=1.08\times10^{18}$m where the plasma is in equipartition.  The population then suffers radiative losses as it travels away from the shock though less severely than at the base of the jet.  

\subsection{Fit 2 - A transition region within the dusty torus}

\begin{figure*}
	\centering
		\includegraphics[width=15 cm, clip=true, trim=1cm 1cm 0cm 1cm]{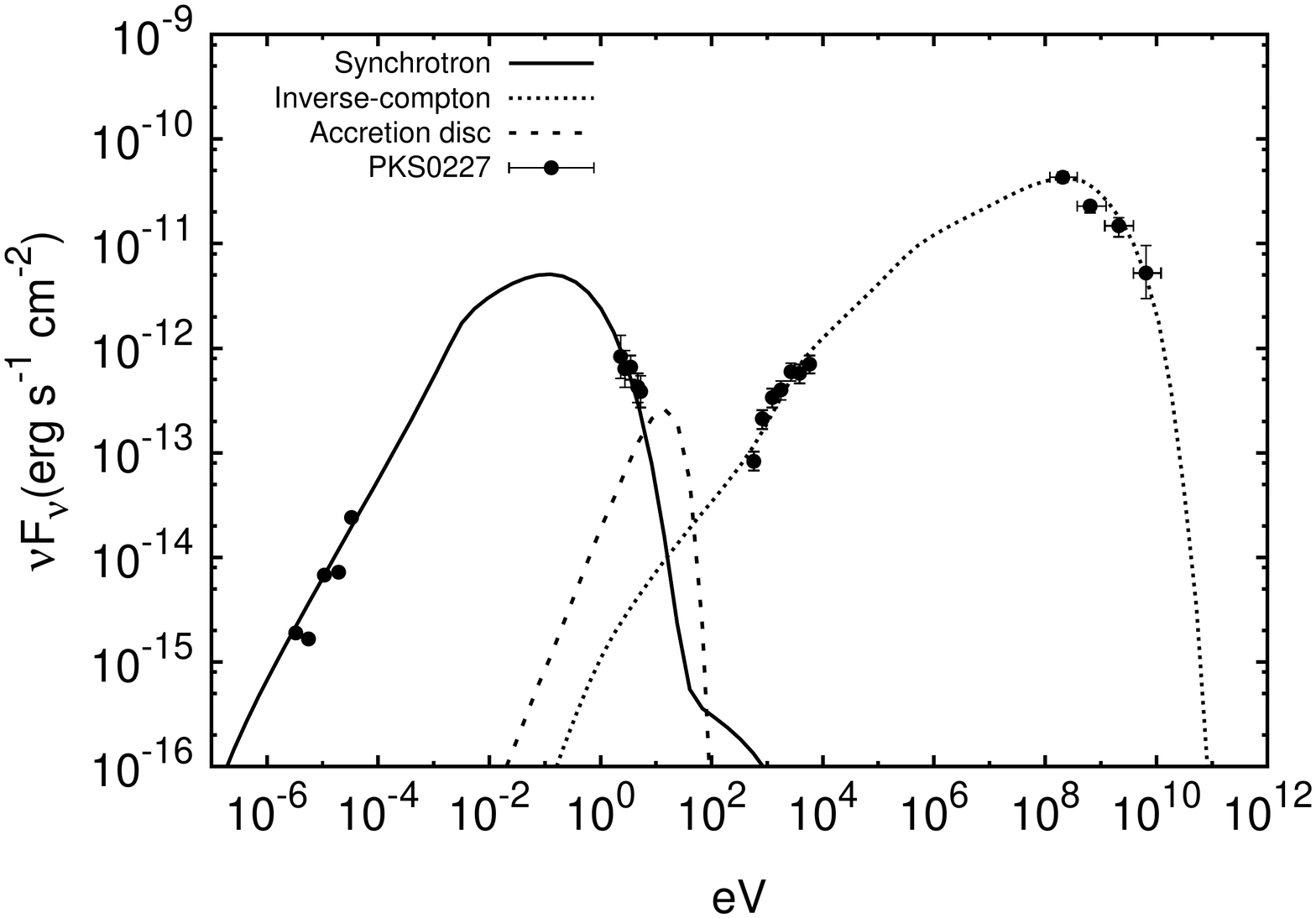}
			
	\caption{This figure shows the results of fitting the model to the SED of PKS0227 using a transition region within the dusty torus. The model fits the observations well across all wavelengths including radio observations.  We find the model fit requires a high jet power, high Lorentz factor jet observed at a small angle to the jet axis.  We find that the inverse-Compton peak is due to scattering of Doppler-boosted dusty torus photons at high energies and CMB photons at low energies. }
	\label{fig4}
\end{figure*}

\begin{figure*}
	\centering
		\subfloat[The fit to PKS0227 showing the different contributions to the inverse-Compton emission.]{ \includegraphics[width=8 cm, clip=true, trim=1cm 1cm 0cm 1cm]{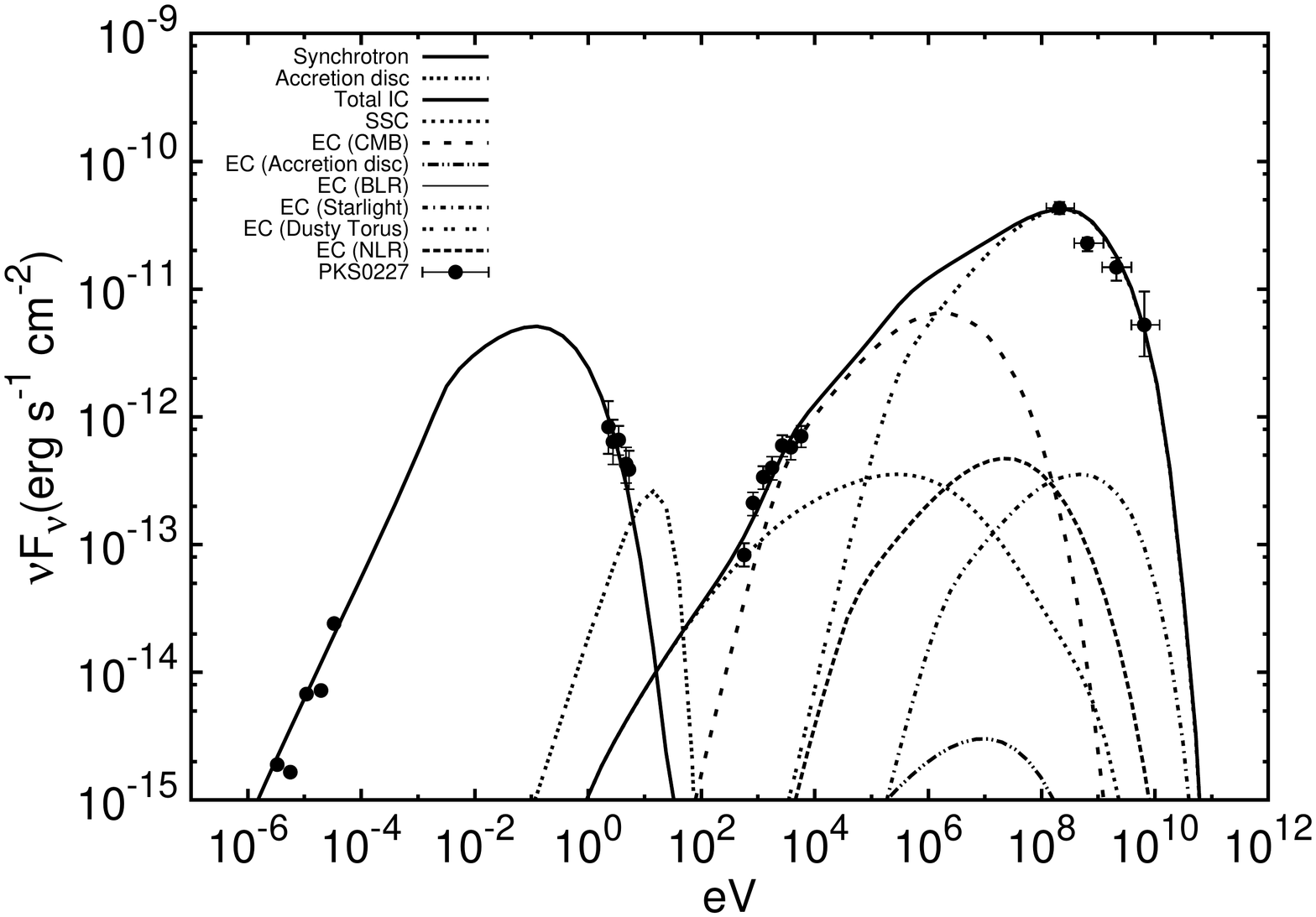} } 
		\qquad
		\subfloat[The fit to PKS0227 showing the emission from different distances along the jet]{ \includegraphics[width=8 cm, clip=true, trim=1cm 1cm 0cm 1cm]{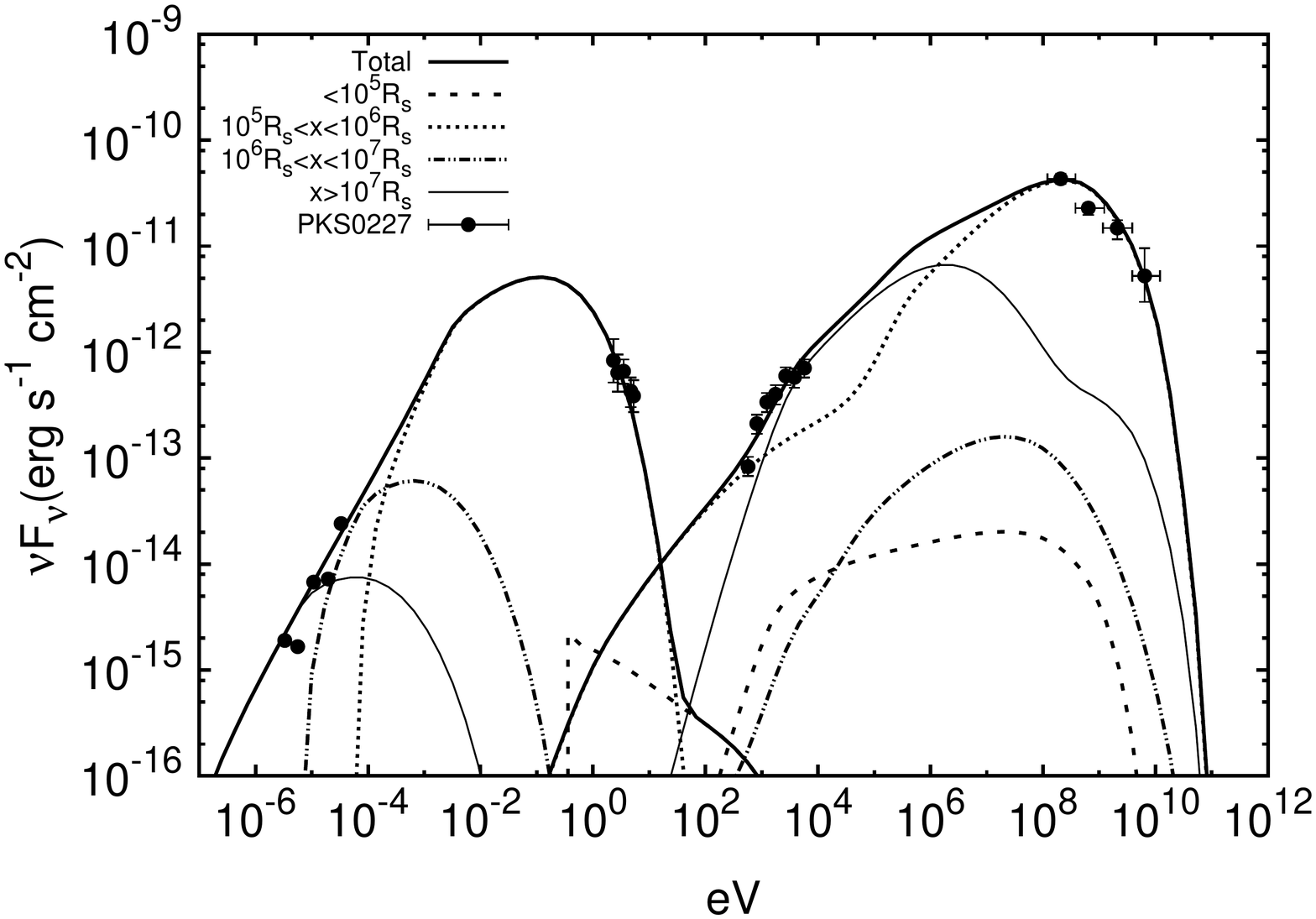} }
	
	\caption{This figure shows the different inverse-Compton components in Figure a and the jet emission from different distances in Figure b, using a transition region within the dusty torus (Fit 2).  Figure a shows that the inverse-Compton emission is produced by scattering dusty torus photons at high energies whilst the x-ray emission is due to scattering CMB photons.  Figure b shows that the optically thin synchrotron emission is dominated by the transition region, with the radio emission coming from larger distances as we expect.  The high energy inverse-Compton emission originates close to the transition region within the torus whilst the x-ray emission due to scattering CMB photons comes from further along the jet.}
	\label{fig5}
\end{figure*}

We now try to fit the spectrum using a transition region within the dusty torus but outside of the BLR.  An emission region within the dusty torus has been investigated previously using one and two-zone models by \cite{2000ApJ...545..107B}, \cite{2002A&A...386..415A},  \cite{2009ApJ...704...38S} and \cite{2011A&A...534A..86T}.

In Figure \ref{fig4} we show the results of our model fitted by eye to the simultaneous multi-wavelength spectrum of PKS0227.  The model fits well to the spectrum across all wavelengths with a transition region located at a distance $1.5$pc from the central black hole with a jet radius of $9.0 \times 10^{14}$m.  A transition region at this distance corresponds to a black hole mass of $1.5\times 10^{8}M_{\odot}$ if the jet has the same geometry as M87 scaled linearly with black hole mass.  In Table \ref{tab1} we show the parameters used in the fit.  Most of the jet parameters are very close to the values found for the fit at large distance in the previous section.  Again the fit requires a high power ($W_{j}=1.5\times 10^{39}$W), high bulk Lorentz factor jet ($\gamma_{\m{max}}=48$) observed close to the jet axis $\theta_{\m{obs}}=0.8^{o}$.  The maximum bulk Lorentz factor is 48, which decelerates to 38 by $x=10^{21}$m.  

In Figure \ref{fig5}a we show the different components of the inverse-Compton emission for the fit.  We see that the dusty torus photons produce the highest energy gamma-rays with the CMB responsible for the low energy x-rays as in fit 1.  Due to the relatively high magnetic field strength in the transition region, $4.81\times10^{5}T$, the jet produces SSC emission which contributes to the lowest energy x-ray observations.  In Figure \ref{fig5}b we show the emission from different distances along the jet.  The synchrotron emission is dominated by the transition region as we expect from the discussion in section 8, with the radio emission coming from larger distances along the jet.  The high energy inverse-Compton emission is dominated by the transition region within the dusty torus, whilst the x-ray emission originates from scattering CMB photons at large distances.

\subsection{Fit 3 - A transition region within the BLR}

\begin{figure*}
	\centering
		\includegraphics[width=15 cm, clip=true, trim=1cm 1cm 0cm 1cm]{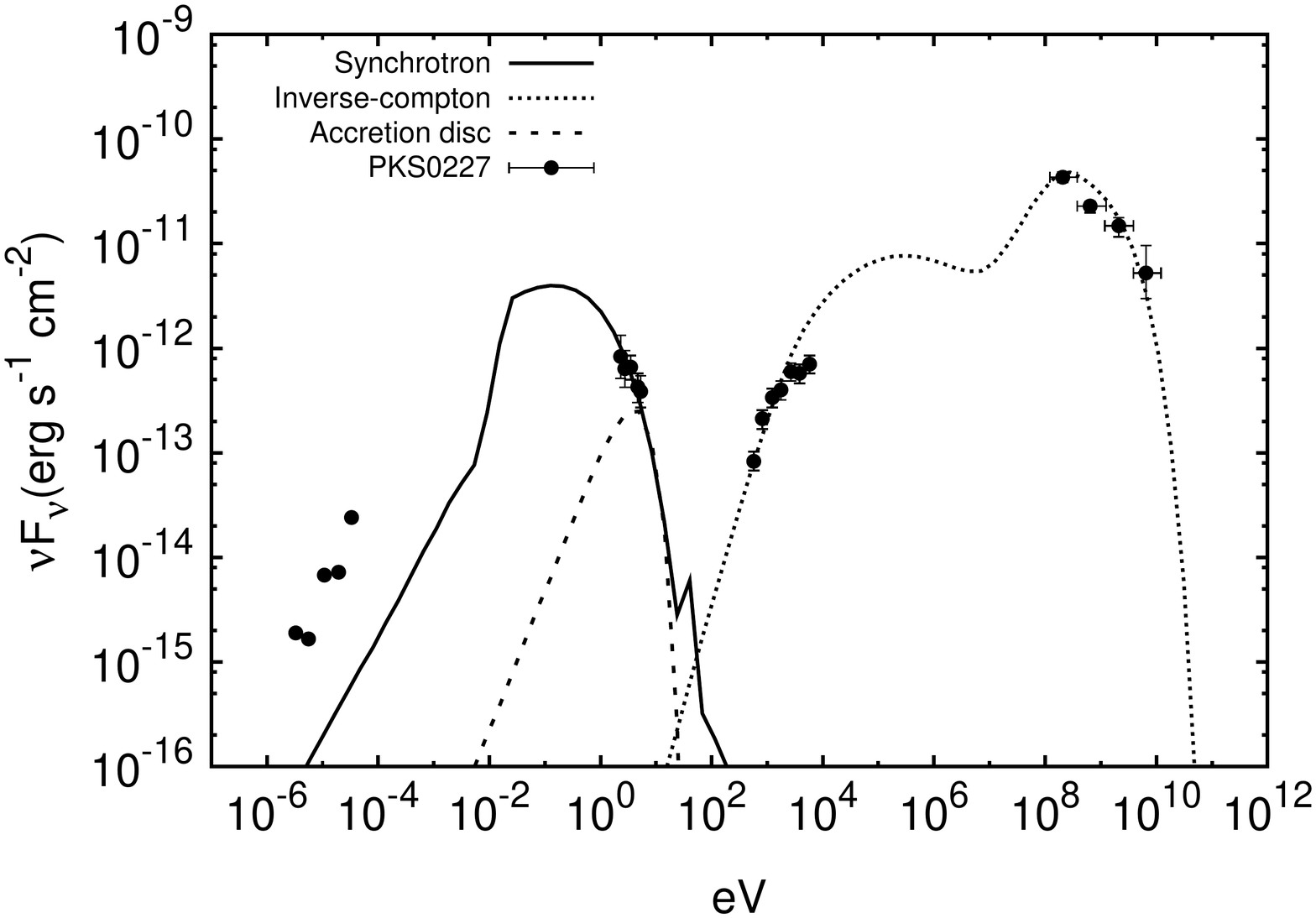}
			
	\caption{This figure shows the results of fitting the model to the SED of PKS0227 using a transition region within the BLR. The model fits the high energy synchrotron and gamma-ray observations reasonably well but is unable to reproduce the radio observations due to the short radiative lifetime of electrons within the BLR.  The fit requires a low maximum electron Lorentz factor to match the gamma-ray peak frequency and a high minimum electron Lorentz factor in order to not overproduce the observed x-ray emission with SSC.  We find that the inverse-Compton peak is due to scattering of Doppler-boosted BLR photons at high energies and SSC photons at low energies. }
	\label{fig6}
\end{figure*}

\begin{figure*}
	\centering
		\subfloat[The fit to PKS0227 showing the different contributions to the inverse-Compton emission.]{ \includegraphics[width=8 cm, clip=true, trim=1cm 1cm 0cm 1cm]{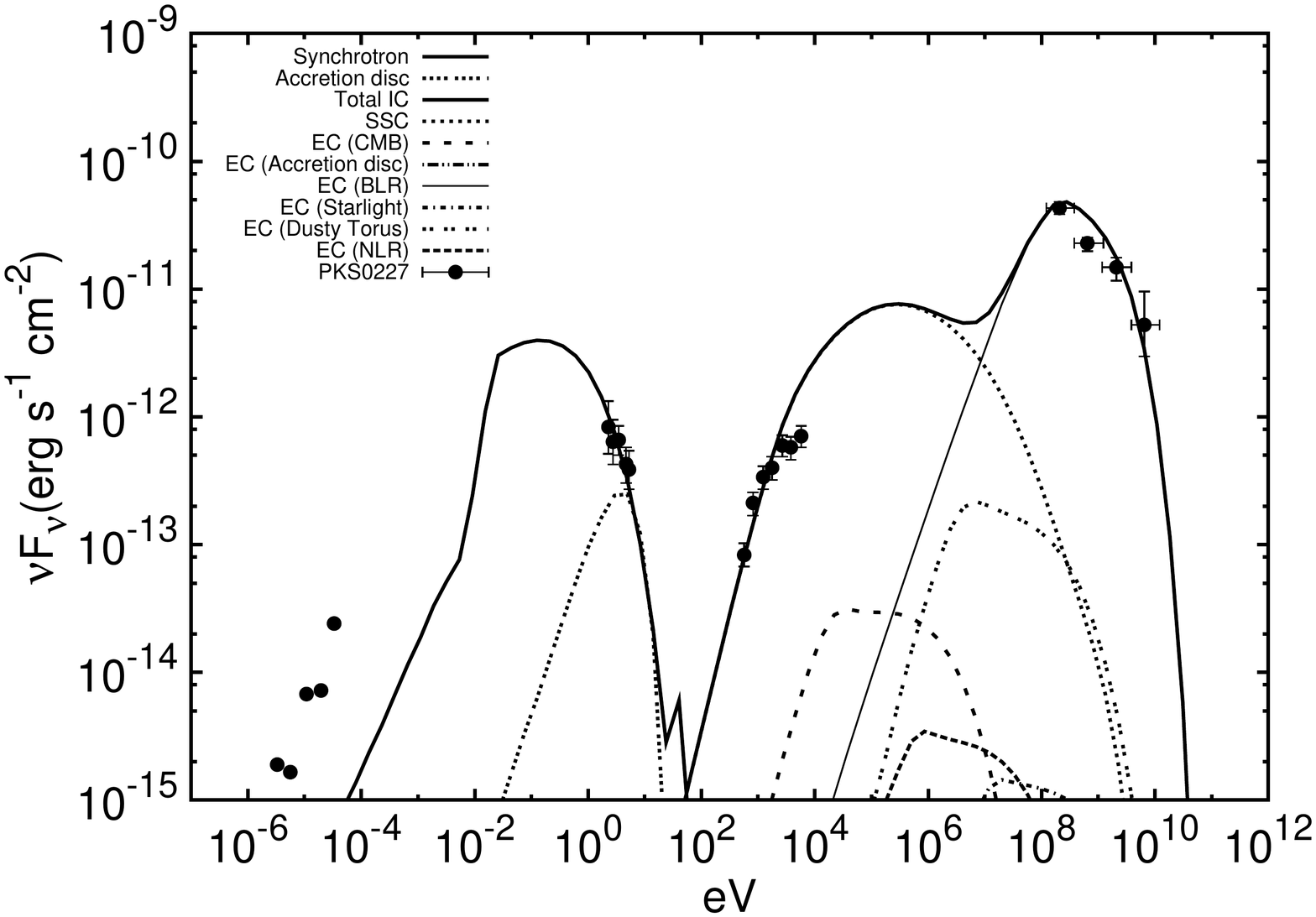} } 
		\qquad
		\subfloat[The fit to PKS0227 showing the emission from different distances along the jet]{ \includegraphics[width=8 cm, clip=true, trim=1cm 1cm 0cm 1cm]{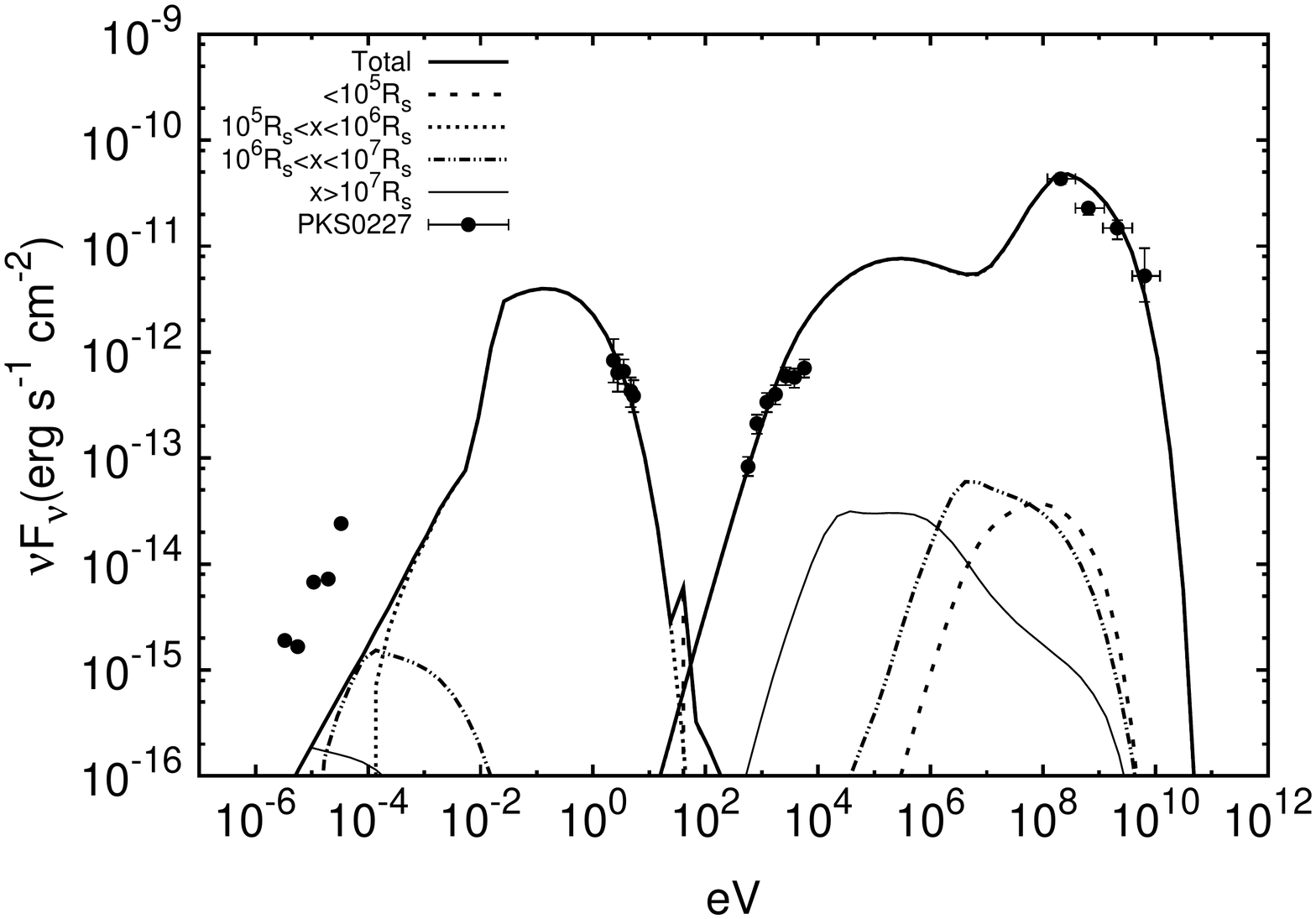} }
	
	\caption{This figure shows the different inverse-Compton components in Figure a and the jet emission from different distances in Figure b, using a transition region within the BLR (Fit 3).  Figure a shows that the inverse-Compton emission is produced by scattering BLR photons at high energies whilst the x-ray emission is due to SSC.  Figure b shows that the optically thin synchrotron emission is dominated by the transition region, with an absence of radio emission from larger distances.  Both the high and low energy inverse-Compton emission originate from close to the transition region within the BLR due to the high magnetic field strength.}
	\label{fig7}
\end{figure*}

Finally we investigate modelling the spectrum using a transition region within the BLR.  An emission region within the BLR has been investigated by many authors including \cite{1994ApJ...421..153S}, \cite{2000AJ....119..469B}, \cite{MNR:MNR13360} and \cite{2009ApJ...692...32D}, and is the most popular location for the gamma-ray emission in blazars due to the ability to reproduce short flaring timescales. 

In Figure \ref{fig6} we show a fit to the spectrum of PKS0227 using a transition region within the BLR and dusty torus.  We show the parameters fitted by eye to the observations in Table \ref{tab1}.  The transition region is located at a distance $0.45\m{pc}$ from the black hole with a jet radius of $2.7\times 10^{14}$m.  This corresponds to a black hole mass of $3.0 \times10^{7}M_{\odot}$ if the jet has the same geometry of M87 scaled linearly with black hole mass.  We see that whilst the model is able to reproduce the high energy gamma-rays and optical synchrotron it is unable to fit the radio or x-ray observations.  We find that due to the short radiative lifetime of electrons in the BLR most of the jet power is quickly radiated away leaving insufficient energy in the jet at larger distances outside the BLR to reproduce the radio emission (see Figure \ref{fig7}b).  This is because the energy in particles is being depleted quickly and in order to maintain equipartition the remaining magnetic energy is constantly converted into replenishing the electrons, resulting in most of the jet power being radiated away. 

The high energy inverse-Compton emission is produced by scattering BLR photons with a low maximum electron Lorentz factor of $851$ (see Figure \ref{fig7}a).  The high magnetic field strength at the transition region due to the small jet radius, $4.59\times 10^{-4}$T, is required to reproduce the high energy synchrotron peak for this low maximum electron energy.  However, this high magnetic field also results in a large amount of SSC emission at x-ray energies which in turn requires a very high minimum electron Lorentz factor of 200 to fit to the x-ray data and not overproduce the observed x-ray emission.  This very narrow range of electron energies is also difficult to understand in the context of shock acceleration where electrons are accelerated from low to high energies in a power law.  

The short radiative lifetime within the BLR due to the high magnetic field strength and large external photon energy density means that any energy contained in high-energy electrons in the jet is quickly radiated away.  This means that either the jet is still magnetically dominated at these distances and that only occasional transient bursts of intense particle acceleration occur within the BLR (possibly causing the observed short-timescale gamma-ray flares) so that the majority of the jet\rq{}s energy is carried to larger distances as required to reproduce radio observations and by the energy requirements of radio lobes.    

We find that the inferred black hole mass of a transition region within the BLR is relatively low and the high peak frequency of the corresponding accretion disc radiation makes fitting the spectrum difficult.  In this fit we use a separate black hole mass for the accretion disc $M_{\m{acc}}$ and jet geometry $M$. For the value of $M_{\m{acc}}$ we use $2 \times 10^{9}M_{\odot}$ found by  \cite{2009MNRAS.399.2041G} from fitting an accretion disc to the spectrum.

\subsection{Constraining the location of the emission region}

We find that both a transition region at large distances and a transition region within the dusty torus are able to reproduce the spectrum of PKS0227 well.  However, we find that a transition region (where the jet comes into equipartition in our model and starts emitting) within the BLR is unable to reproduce the observations due to the short radiative lifetimes of electrons within the BLR.  We find that the jet must still be magnetically dominated within the BLR since the plasma is unable to maintain a state close to equipartition due to high radiative energy losses which quickly drain the jet of energy.  This means it is unable to produce sufficient radio emission at large distances and is also incompatible with the proportion of jet energy deposited in radio lobes.  It is possible that occasional bursts of particle acceleration occur within the BLR (which could be responsible for short timescale gamma-ray flares).

We find that both fits to the spectrum of PKS0227 using a transition region at large distances and within the dusty torus require large jet power, high bulk Lorentz factor jets observed close to the line of sight.  This is consistent with Compton-dominant sources being FRII type jets.  For both these fits we find that the x-ray emission is due to inverse-Compton scattering of the CMB at large distances.  This is the first time that the CMB has been investigated in a model for blazar emission so this is an interesting result.

The black hole mass corresponding to a transition region located at $10^{5}R_{s}$ as in M87 is $3.4\times10^{9}M_{\odot}$ and $1.5\times10^{8}M_{\odot}$ for fits 1 and 2 respectively.  A previous investigation by \cite{2009MNRAS.399.2041G} found a black hole mass of $2\times 10^{9} M_{\odot}$ by modelling an accretion disc spectrum to PKS0227.  This black hole mass is close to the value of fit 1 which uses a transition region at large distances but is substantially larger than fit 2 with a transition region inside the dusty torus. 

Whilst our mass estimate of the black hole inferred from fit 1 is close to that found by  \cite{2009MNRAS.399.2041G} the values of our jet parameters differ considerably from theirs.  This is due to fundamentally different approaches to modelling the jet.  In their model they consider a single compact blob which produces the observed inverse-Compton emission from scattering BLR photons whilst the optical observations are matched using thermal accretion disc, torus and corona components.  Since they only consider a single emitting region they do not attempt to reproduce the radio data.  In this model they find a bulk Lorentz factor of 14 substantially smaller than in our jet, a viewing angle is $3^{o}$ slightly larger than in our fit and a total jet power is higher by over an order of magnitude $2.38 \times 10^{40}W$, however, their model includes one proton per electron.  These very different parameters are due to fundamentally different model assumptions.  It is worth pointing out that no attempt is made to fit the model to the observed radio data and in addition the synchrotron radiation also does not fit the optical data since this is fitted by the thermal components, so the non-thermal synchrotron emission is not well constrained.

The geometry of M87 suggests that the transition region where the jet plasma approaches equipartition is at relatively large distances from the black hole, $>10\m{pc}$ for black holes of mass $>10^{9}M_{\odot}$.  This puts the transition region outside the BLR and outside or near to the outer edge of the dusty torus so we do not expect photons from these regions to dominate the external radiation field at these distances (see Figure \ref{fig3}).  We find that this is consistent with our model of the observed spectrum of PKS0227 which is well fitted by a transition region at a distance $34\m{pc}$ from the black hole where the inverse-Compton emission is due primarily to scattering of CMB photons.  The transition region may not occur at $10^{5}R_{s}$ for every jet as it could depend on the jet power, mass loading etc.  If this were the case we might expect a jet with a high bulk Lorentz factor to have a transition region at a larger distance than the comparatively low power jet in M87, in order to accelerate to a higher bulk Lorentz factor.  This is not the case for fit 2 where the transition region would have to be located at $7500R_{s}$ to be compatible with the black hole mass estimate of \cite{2009MNRAS.399.2041G}.  For these reasons we favour a scenario in which the transition region occurs at large distances outside the BLR and dusty torus.

From the Figures \ref{fig1} and \ref{fig4} we see that one of the distinguishing features between a transition region at large distances or within the dusty torus is the location of the optically thick to thin synchrotron break.  Since the optically thin synchrotron emission is dominated by the transition region (see Figures \ref{fig2}b and \ref{fig5}b) the location of the transition region largely determines the frequency at which the optically thick to thin break occurs.  We see that the frequency at which the flat self-absorbed radio spectrum becomes optically thin and changes gradient increases for a transition region at a smaller distance along the jet as we expect due to the increased magnetic field and smaller jet radius.  This feature could potentially be used to distinguish between a transition region at large distances or within the dusty torus.  We wish to investigate whether this can be used to constrain the emission region in the near future.

\subsection{An adiabatic or ballistic conical jet?}

\begin{figure*}
	\centering
		\subfloat[Fit 1 to PKS0227 with and without adiabatic losses in the conical part of the jet.]{ \includegraphics[width=8 cm, clip=true, trim=1cm 1cm 0cm 1cm]{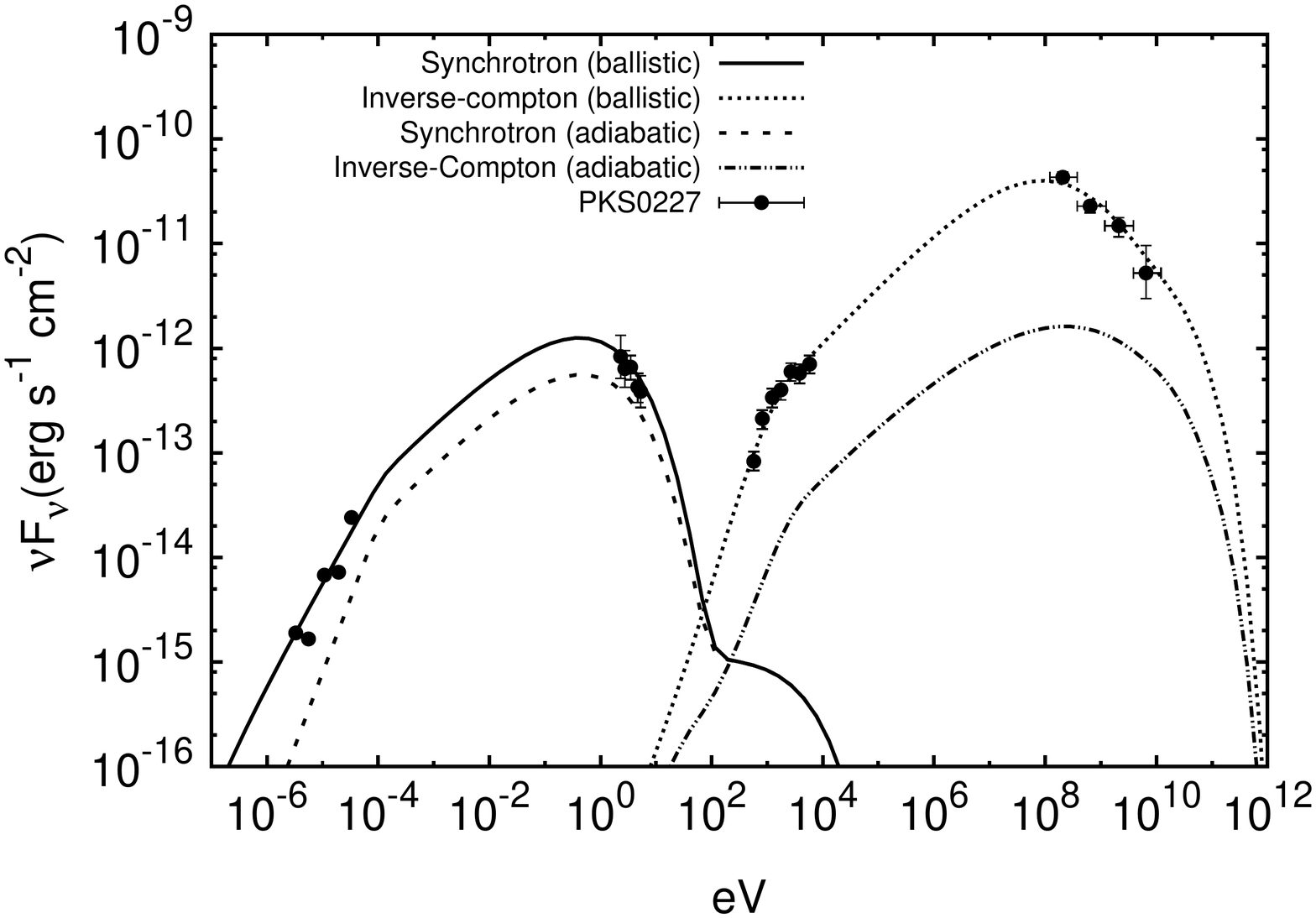} } 
		\qquad
		\subfloat[Fit 2 to PKS0227 with and without adiabatic losses in the conical part of the jet.]{ \includegraphics[width=8 cm, clip=true, trim=1cm 1cm 0cm 1cm]{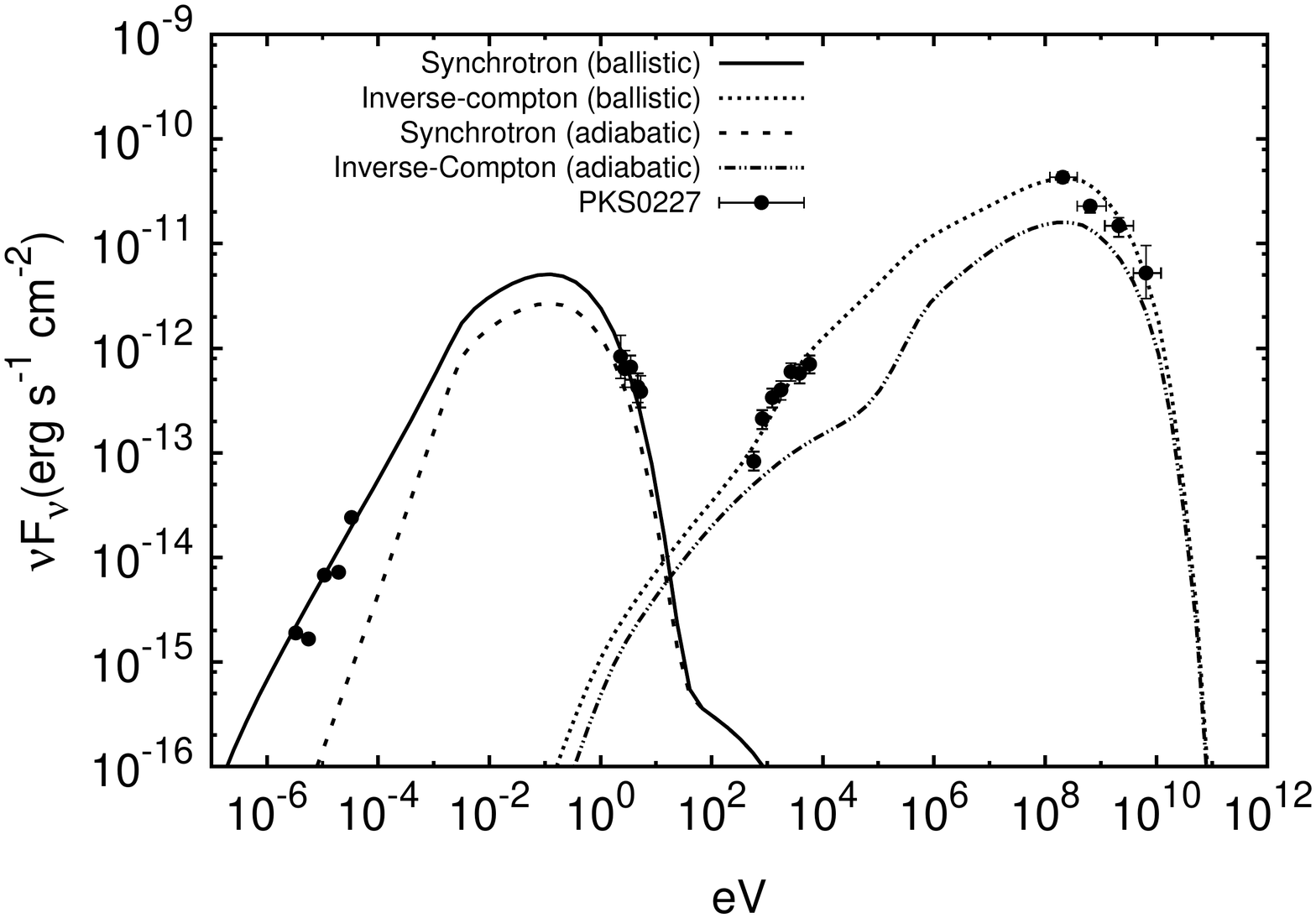} }
	
	\caption{This figure shows fits 1 and 2 to the spectrum of PKS0227 with and without adiabatic losses in the conical section of the jet.  The Figure shows that the result of adiabatic losses on the electron population is to severely reduce the amount of inverse-Compton radiation, especially from scattering CMB photons at large distances.  The radio emission is reduced as the jet loses additional energy to adiabatic losses at large distances where the radio emission originates.  We find it difficult to reproduce the spectrum using a transition region at large distances with adiabatic losses without using a large bulk Lorentz factor.  We also find it difficult to reproduce the spectrum using a transition region within the dusty torus due to the reduction in x-ray and radio emission.}
	\label{fig8}
\end{figure*}

In this paper we have assumed that the parabolic accelerating region of the jet expands adiabatically whilst the conical region is ballistic.  In Paper I we proposed that a continuous conical jet with a constant opening angle and bulk Lorentz factor would not suffer adiabatic losses because it does no work expanding its environment (since the volume occupied by the jet is constant with time in the lab frame).  In an idealised homogeneous jet the velocity profile in the direction perpendicular to the jet axis scales linearly with expansion and so small sections of the jet maintain a constant perpendicular velocity and are effectively ballistic.  Whilst these arguments are based on an idealised situation, a conical jet with an opening angle $\theta_{\m{opening}}\gamma_{\m{bulk}}>1$ is not causally connected laterally and analytic work by \cite{2008ApJ...679..990Z} found that far away from the black hole a conical jet will always become causally disconnected and this seems to be the case for jet parameters inferred from observations of GRBs \cite{2012arXiv1208.1261K}.  Work reproducing the large scale observed emission and polarisation of FRI jets using MHD models by \cite{2004MNRAS.348.1459L} found that jets were better described by non-adiabatic jets as opposed to adiabatic jets.  It is therefore worth testing the effect of including adiabatic losses in the conical section of our model.

In Figure \ref{fig8} we show the results of including adiabatic losses on fits 1 and 2.  We find that in fit 1 adiabatic losses cool the high energy electrons more quickly than scattering the CMB and so the inverse-Compton emission is vastly reduced.  In the case of adiabatic losses we find it difficult to fit the spectrum of PKS0227 with a transition region at large distances without using very large bulk Lorentz factors.  In the case of fit 2 we also find that adiabatic losses substantially decrease the inverse-Compton emission from the jet especially the x-ray emission which was due to scattering CMB photons.  We also find that adiabatic losses result in a decrease in the radio emission at large distances due to the adiabatic energy losses along the jet.  We find it difficult to fit the spectrum of PKS0227 at radio and x-ray energies using a transition region within the dusty torus including adiabatic losses.  

For these reasons we favour a ballistic conical section of the jet.  This could be a result of the assumptions we have made when fitting the spectrum with our jet model.  We have assumed that the jet comes into equipartition at the transition from parabolic to conical and the jet maintains equipartition outwards from this region.  This is motivated by simulations which indicate the jet transitions to conical when the plasma approaches equipartition and from observations which seem to indicate that the jet plasma is observed to be close to equipartition with fairly continuous optical emission requiring continuous particle acceleration along the jet.  In this paper we have assumed a simple scenario in which the jet suddenly transitions to equipartition, however, it is likely that some particle acceleration and emission comes from the magnetically dominated region of the jet and this could affect our conclusions.  We intend to investigate different mechanisms for particle acceleration along the jet in the near future.

We have found that the x-ray emission is well fitted by inverse-Compton scattering of CMB photons at large distances along the jet.  This x-ray emission is severely reduced if the conical section is adiabatic and becomes difficult to reproduce by non-thermal emission from the jet.  The x-ray emission may also be due to the accretion disc corona so an inability to reproduce the x-ray emission with the jet does not necessarily exclude an adiabatic conical jet.  We are keen to investigate whether this result holds for other Compton-dominant jets in the next paper in the series. 

\subsection{Discussion}

In this work we have attempted to model the quiescent state of the blazar PKS0227 since this constrains the average jet parameters.  Short term variability is observed on time-scales as short as hours or minutes for the most highly variable blazars (\cite{2007ApJ...664L..71A}).  The quiescent emission from our jet model comes from large distances with long electron cooling times ( $\sim$years for fit 1 and $\sim$days for fit 2), so is not easily compatible with this short timescale variability.  In this series of papers we wish to initially investigate the average quiescent emission from different elements of the blazar population in order to check that our model is able to reproduce their spectra and to find any trends in the quiescent jet parameters.  Once we are confident that our model is able to reproduce the quiescent spectra of both BL Lacs and FSRQs with reasonable jet parameters, we will investigate whether we can reproduce flaring spectra by allowing flares to occur within the parabolic region of the jet.  The parabolic region of the jet is more compact with shorter cooling timescales due to the higher magnetic field strength and energy density of external photons, so this region is more likely to be responsible for short timescale flaring.    

It is important to emphasise that whilst our model is more complicated than a single zone model it does not possess many more free parameters.  We have attempted to minimise the number of free parameters in our model by making reasonable, simplifying physical assumptions and using the results of previous work observing and modelling the external radiation fields.  The CMB is well constrained by observations and possesses no free parameters, however, the emission and distribution of the NLR is less well known.  We have fixed most of the free parameters using values from previous observations and modelling of the BLR, dusty torus and NLR and we have left only the outer radius $x_{\m{outer}}$ of the NLR as a free parameter since this seems to vary the most significantly.  In these fits we have not injected an arbitrary electron population into arbitrary sections along the jet, we have injected a small amount of the initial jet power ($1\%$) into electrons at the very base of the jet (to illustrate the resulting spectrum from electrons at the base of the jet whilst not affecting the fit to observed data, see Figure \ref{fig2}b).  The acceleration of electrons in the decelerating section of the jet is determined entirely by Equation \ref{cont1} in terms the other jet parameters in Table \ref{tab1} and assuming equipartition in the conical section.  Along with the fixed geometry this leaves our model with twelve free parameters which we vary when fitting the spectrum ($W_{j}$, $L$, $L_{acc}$, $\theta_{\m{obs}}$, $\theta_{\m{opening}}$, $M$, $\alpha$, $E_{\m{min}}$, $E_{\m{max}}$, $\gamma_{\m{\max}}$, $\gamma_{\m{min}}$ and $x_{\m{outer}}$).  This number of free parameters is comparable to single-zone models with broken power-law electron distributions.

We have tested the effect of changing the functional form of the deceleration of the bulk Lorentz factor along the jet (Equation \ref{dec}).  In our model the bulk kinetic energy of the jet is spread evenly on a logscale along the jet from deceleration.  We find that strongly decelerating the jet over a particular distance scale effectively boosts the relative amount of emission from that section of the jet.  This means that if the jet decelerates strongly over a scale $\sim10^{5}R_{s}$ then the amount of optical synchrotron and SSC emission is boosted relative to the radio emission and inverse-Compton scattering of CMB photons.  Similarly if the jet is chosen to decelerate strongly at larger scales the amount of low frequency radio and inverse-Compton emission increases relative to the optical synchrotron and SSC emission.  We find that strong deceleration on an intermediate length scale leads to a bump in the characteristic radio frequencies of that section.  This deviation from a flat radio spectrum is not something seen in radio observations of blazars and so we disfavour strong deceleration of the jet on an intermediate length scale $10^{17}-10^{19}m$.  We find that deceleration which decreases as the logarithm of the jet distance as we have assumed (Equation \ref{dec}) distributes power more evenly along the jet resulting in a radio spectrum which is smooth and close to flat consistent with observations. 

We find that electrons injected into the parabolic section of the jet quickly radiate all their energy away due to the large magnetic fields and high energy density in the seed photon field (see Figure \ref{fig7}b).  We also find that the emission of these electrons tends to be dominated by the emission from the conical section of the jet, since the inner regions of the accelerating parabolic section have a lower bulk Lorentz factor and so are not so strongly Doppler-boosted.  Since all the energy injected into electrons in the inner parts of the parabolic region is quickly radiated away and does not contribute as significantly to the observed spectrum it is difficult to constrain the amount of energy injected into accelerating electrons in this region.  Since we expect the jet to be magnetically dominated in the parabolic region we have assumed that a small percentage of the magnetic energy goes into shock acceleration of the electrons at the base of the parabolic section, this means that our estimated jet power is a lower bound because the total jet power would increase if a larger proportion of magnetic energy were converted into particle energy in the inner parabolic region, without a significant increase in observed emission.

We find that the low frequency radio emission of the jet is dominated by the conical sections with the lowest bulk Lorentz factors and largest radii (see Figure \ref{fig2}b).  This is partly because the regions with high bulk Lorentz factors are strongly Doppler-boosted so the synchrotron opacity in these sections is effectively the synchrotron opacity for a much lower energy radio photon than that observed.  The synchrotron opacity increases steeply at low frequencies so the radio emission from regions with high bulk Lorentz factors is suppressed and the strongly decelerated regions with larger radii dominate the radio.  We see that the model is able to fit to the nearly flat radio spectrum of PKS0227 very well.  Including the effects of deceleration and the conversion of bulk kinetic energy into particle acceleration means that the electron energy losses are replaced as they travel along the decelerating conical part of the jet as expected from observations of optical synchrotron emission far from the base of a jet (\cite{1997A&A...325...57M} and \cite{2001A&A...373..447J}).  This mechanism to rejuvenate the high energy electron population has been investigated a number of times previously, however, this is the first investigation that has included the effects of deceleration in a continuous conical jet which conserves energy-momentum and particle number.

Most previous investigations have neglected radio observations and concentrated on one or two-zone models which are disconnected from the large scale radio emission of the jet.  We have demonstrated that our rigorous, physically motivated, inhomogeneous jet model fits well to the simultaneous multiwavelength spectrum of PKS0227 across all wavelengths including radio observations, which are usually ignored.  This allows us to better constrain the regions where emission may occur in the jet by reproducing the radio and large scale emission of the jet whilst conserving energy along the jet.  We find the results of this initial application of our model to a blazar very promising and we intend to use this model to conduct a more detailed study of a sample of Compton-dominant blazars in the near future to see if our results hold for other blazars and to try to distinguish between a transition region at large distances or within the dusty torus.

\section{Conclusion}
 
In this paper we have developed a realistic jet model which includes an accelerating, magnetically dominated, parabolic base which transitions to a slowly decelerating conical jet with a jet geometry set by recent radio observations of M87 and consistent with simulations and theory.  This is the first time that an accelerating parabolic region transitioning to a decelerating conical region has been treated in a jet model for the spectra of a blazar.  We calculate the observed line of sight synchrotron opacity to each section along the jet by Lorentz transforming into the rest frame of the plasma.  We calculate the inverse-Compton emission from accretion disc photons directly from the disc, BLR, dusty torus, NLR, starlight and CMB photons by Lorentz transforming the radiation into the plasma rest frame.  Our model conserves relativistic energy-momentum and particle number along the jet and we take into account radiative and adiabatic losses on the electron population.  

We assume that the jet is initially magnetically dominated until it transitions from parabolic to conical at $10^{5}R_{s}$, where the jet stops accelerating and the plasma approaches equipartition.  Electrons are accelerated at this transition region and electron energy losses are replenished along the jet by the conversion of bulk kinetic energy into particle energy as the jet decelerates so that the plasma is kept in equipartition throughout the conical section of the jet.  We use the model with a geometry and acceleration set by VLBI observations of M87 and simulations to fit to the simultaneous multi-wavelength spectrum of the Compton-dominant FSRQ PKS0227.  

We investigate a transition region located outside the dusty torus, within the dusty torus and within the BLR.  We find that our model fits very well to the spectrum across all wavelengths including radio observations with a transition region at either large distances from the central black hole, or within the dusty torus.  We find that due to the short radiative lifetime we are unable to reproduce the radio observations using a transition region within the BLR.  We find that within the BLR the jet must still be magnetically dominated in order for the jet to transport sufficient energy to larger distances, however, this does not exclude occasional transient bursts of intense particle acceleration.   

We find that our fits using transition regions outside the BLR both require high power ($\sim10^{39}$W), high bulk Lorentz factor jets ($>40$) viewed at a small angle to the jet ($\sim1^{o}$).  We find an inferred black hole mass of $3.4\times 10^{9}M_{\odot}$ for the transition region outside the dusty torus and $1.5 \times10^{8}M_{\odot}$ for a transition region within the dusty torus using observations of the geometry of M87 scaled linearly with black hole mass.  The inferred black hole mass of the transition region at large distances, which completely determines the inner geometry of the jet, is close to the black hole mass found by \cite{2009MNRAS.399.2041G} ($2\times 10^{9}M_{\m{\odot}}$).  We find that the optically-thick to thin synchrotron break depends on the location of the transition region and can potentially be used to distinguish between the two emission locations.

Interestingly, we find that including the redshift dependence of the CMB and by boosting into the plasma rest frame the Compton-dominance of the transition region at large distances $34$pc is well fitted by inverse-Compton scattering of CMB photons.  This could explain why Compton-dominant blazars are powerful, high bulk Lorentz factor blazars which are preferentially at high redshift due to the temperature dependence of the CMB $\propto (1+z)$.  We find that the x-ray emission for both transition region fits outside the BLR can be fitted well by scattering of CMB photons on large scales if the conical section of the jet is ballistic.

We find these initial results very promising and we intend to use this model as the basis for a more detailed investigation into the physics behind different elements of the blazar population in the near future.

\section*{Acknowledgements}

WJP acknowledges an STFC research studentship. GC acknowledges support from STFC rolling grant ST/H002456/1.  We would like to thank Jim Hinton, Brian Reville and Tony Bell for useful discussions about the model.  WJP acknowledges an STFC LTA grant which allowed him to visit Stanford University where he had valuable discussions with Roger Blandford, Jonathan Mckinney and Greg Madejski.   

\bibliographystyle{mn2e}
\bibliography{Jetpaper2refs}
\bibdata{Jetpaper2refs}

\label{lastpage}

\end{document}